\documentclass[11pt]{article}
\usepackage{epstopdf}
\usepackage[a4paper,hmarginratio=1:1,vmarginratio=2:3,totalwidth=15.3cm,totalheight=22.7cm]{geometry}
\usepackage{bm,epstopdf,epsfig,amsmath,amssymb,amsfonts,colordvi,latexsym,comment,cancel,verbatim,
slashed}
\usepackage{graphicx,graphics}
\usepackage[font=md,captionskip=8pt]{subfig}
\usepackage[usenames,dvipsnames]{color}
\usepackage[noadjust]{cite}

\usepackage{xcolor}  

\newcommand{\m}{\,-\,}
\newcommand{\p}{\,+\,}

\usepackage{setspace}

\usepackage[utf8]{inputenc}
\allowdisplaybreaks

\newcommand{\cO}{{\cal O}}

\newcommand{\cV}{{\mathcal V}}

\newcommand{\ra}{\rightarrow}
\newcommand{\be}{\begin{equation}}
\newcommand{\ee}{\end{equation}}
\newcommand{\bea}{\begin{eqnarray}}
\newcommand{\eea}{\end{eqnarray}}

\newcommand{\baa}{\begin{array}}
\newcommand{\eaa}{\end{array}}
\long\def\symbolfootnote[#1]#2{\begingroup
\def\thefootnote{\fnsymbol{footnote}}\footnote[#1]{#2}\endgroup}
\begin{document} 
\begin{flushright}
\end{flushright}

\bigskip\medskip

\thispagestyle{empty}

\vspace{3.cm}

\begin{center}
 {\Large {\bf   Two-loop corrections to Starobinsky-Higgs inflation}}

\bigskip

\vspace{1.cm}

 {\bf D. M. Ghilencea}
\symbolfootnote[1]{E-mail: dumitru.ghilencea@cern.ch}

\bigskip
{\small Theoretical Physics Department, National Institute of Physics}

{\small and Nuclear Engineering  (IFIN) Bucharest\, 077125, Romania}

\end{center}

\medskip
\begin{abstract}
\noindent
Higgs inflation and $R^2$-inflation (Starobinsky model)  are two limits of the same 
quantum model, hereafter called Starobinsky-Higgs. We analyse the  two-loop 
action of the Higgs-like scalar $\phi$  in the 
presence of: 1) non-minimal coupling ($\xi$) and 2) quadratic  curvature terms. 
The latter are generated at the  quantum level with  $\phi$-dependent couplings 
($\tilde\alpha$) even if their tree-level couplings ($\alpha$) are tuned  to zero.  
Therefore, the potential always depends on both Higgs field $\phi$ and scalaron $\rho$, 
hence multi-field inflation is a quantum  consequence.  The effects of the 
quantum (one- and two-loop) corrections on the  potential   $\hat W(\phi,\rho)$  
and on the spectral index are  discussed, showing that the Starobinsky-Higgs model  is in general
stable in their  presence.
Two  special cases are also considered: first, for a  large $\xi$ in the quantum action one can 
integrate $\phi$   and generate a ``refined''  Starobinsky  model which contains
additional terms  $\xi^2 R^2\ln^p (\xi \vert R\vert/\mu^2)$, $p\!=\!1,2$ 
($\mu$ is the subtraction scale).  These generate  
corrections linear in the scalaron  to the ``usual'' Starobinsky potential  and a 
``running'' scalaron mass. Second, for a small fixed Higgs field $\phi^2\!\ll\! M_p^2/\xi$ and
 a  vanishing  classical  coefficient of the $R^2$-term, we show that 
the  ``usual'' Starobinsky inflation   is generated by the quantum corrections alone,
 for a  suitable non-minimal  coupling ($\xi$).
\end{abstract}

\newpage

\section{Introduction}

The  idea of  inflation in the  early universe
\cite{Guth,Linde:1981mu,Albrecht:1982wi,Starobinsky:1980te,Sato,Linde,Hawking,Hartle} 
(for a review \cite{Sato:2015dga}) 
led to many models in agreement with the cosmic microwave 
background CMB \cite{CMB}; of these, minimal models like  Starobinsky  model 
\cite{Starobinsky:1980te} and the Higgs inflation model \cite{C,S,B}
are among the most successful (for more recent developments in 
 Higgs inflation see e.g. \cite{Bez1,L1,K1,Bez2,Bez3,E1,H1,A1,G1,B2,B3,SH1,SH2,SH3,SH4,O1} and in
$R^2$-models   \cite{C1,LAG,AS,K2,C2,Giudice,N2,M1,W1,G2,B4,G3,C4}).

 In Higgs inflation  a non-minimal coupling $\xi\phi^2 R$ of the Higgs $\phi$ 
 to the Ricci scalar $R$  is considered, with $\phi$ also in the role of  the inflaton.
 This relates cosmology to Standard Model precision tests. 
 In  Starobinsky inflation an  $\alpha R^2$ term ($\alpha$ constant)
  is added to the  Einstein term, thus  inducing geometrically a  new scalar field  $\rho$
 (scalaron) playing the role of inflaton. Both  models give a similar spectral index 
 of primordial scalar  adiabatic  perturbations.

If quantum corrections of matter are included,  these two models are  special limits of a 
single model of inflation (hereafter ``Starobinsky-Higgs''). 
Indeed, consider the quantum corrections in  curved space-time due to 
$\phi$ to the  Higgs potential  $V_0(\phi)$ of flat space-time.
 New terms linear in $R$ e.g. $V_1(\phi)\!\sim \! R\,\phi^2\ln\phi$ 
and quadratic in $R$, Ricci  ($R_{\mu\nu}$) or Riemann  ($R_{\mu\nu\rho\sigma}$) tensors, e.g.
 $V_2(\phi)\!\sim \!\! R^2 \ln \phi$,  emerge in the quantum action
with $\phi$-dependent coefficients,  even if they are  absent at
tree-level. $R^2$-inflation and Higgs inflation are thus unified
in a single quantum  model of fields ($\phi,\rho$), so multi-field inflation 
is  a quantum  consequence.

The  main goal is to study the effects of  quantum corrections of matter to 
the Starobinsky-Higgs model. This study is useful for precision tests of this model.
With either the  Higgs or the $R^2$-term in the action,  one can achieve successful 
inflation.  Their ``overlap'' in the  general parameter region
of Starobinsky-Higgs inflation is then not expected to give dramatic deviations 
from  $R^2$-inflation  or Higgs inflation  alone. One linear combination 
$\phi$-scalaron is essentially responsible for inflation; its quantum 
fluctuations are adiabatic perturbations becoming the seeds of 
inhomogeneities seen in the CMB temperature anisotropy. 
 The other  combination gives rise to isocurvature 
perturbations \cite{sasaki1} specific to  multi-field inflation.

The plan of the paper is as follows. We first review  at the classical level the
Starobinsky-Higgs inflation (Section~\ref{classical}). For special values of the fields and
$\xi$,  $\alpha$ (coefficient of $R^2$) and $\lambda$ (higgs self-coupling), 
one has  Higgs inflation, $R^2$-inflation, or a combination of these.  
We study the two-loop corrections to the effective action  with non-minimal coupling $\xi$
 and quadratic curvature terms ($R^2$, $R_{\mu\nu}^2$, $R_{\mu\nu\rho\sigma}^2$), following
\cite{O1,Odintsov,PT,Eli1,Eli2,Eli3} (Section~\ref{quantum}). The metric is not quantised.
One expands  the potential in powers of curvature \cite{PT}
and solves iteratively  the two-loop Callan-Symanzik equations for $V_0(\phi)$, $V_1(\phi)$,
 $V_2(\phi)$ (Appendix).  We then study the Einstein frame potential $\hat W(\phi,\rho)$
which also depends  on the scalaron.
The effects  of the quantum corrections on the classical potential and on the spectral index $n_s$
are analysed. In general the Starobinsky-Higgs
 model is stable in their presence (Section~\ref{pheno}).

Two special cases are also studied. At large $\xi$, integrating the Higgs field 
in  the quantum  action generates  a ``refined'' Starobinsky model  with  extra  terms 
 $\xi^2\,R^2 \ln^n (\xi \vert R\vert/\mu^2)$, $n\!=\!1,2$; these 
bring terms $\propto\!\rho$  and a running scalaron mass. 
Terms like $C^2\ln\vert R\vert/\mu^2$ and
$G\ln\vert R\vert/\mu^2$ 
 are also generated, with $C^2$ the square of 
Weyl tensor, $G$: Gauss-Bonnet term (Section~\ref{largexi}).
Finally, we show that  even if  there is no $R^2$-term at the tree-level ($\alpha=0$),
 a term $(\xi\!+\! 1/6)^2 R^2 \ln\phi$  emerges at the loop level.
This is an interesting result, since the mere presence of a fixed, small 
Higgs field  $\phi\!\ll\! M_p$ with large non-minimal coupling $\xi\,\phi^2\,R$,   
provides  a  {\it quantum origin} to the   ``usual''  Starobinsky model of inflation!

\section{Starobinsky-Higgs model: classical picture}\label{classical}

Our starting action is that of a scalar theory $\phi^4$ in curved space-time
with quadratic terms\footnote{Our
conventions  \cite{GR}: metric:
 $(+,-,-,-)$; $R^\lambda_{\mu\nu\sigma}=
\m\partial_\sigma\Gamma_{\mu\nu}^\lambda
\p\partial_\nu \Gamma_{\mu\sigma}^\lambda+\cdots$,
 $R_{\mu\nu}=R^\lambda_{\mu\lambda\nu}$, $R=g^{\mu\nu} R_{\mu\nu}$; 
 $R<0$ at inflationary stage; $\xi\!=\!-\frac16$ is conformal.
To use conventions \cite{PT}  do $R\ra -R$, $\xi\ra -\xi$, $\beta_\xi\ra-\beta_\xi$.} 
 \smallskip
\bea\label{L}
 S\!\!&=&\!\!\int d^4 x {\sqrt g}\,\,\Big\{\,
\frac{1}{2} g^{\mu\nu} \partial_\mu\phi\, \partial_\nu\phi
-\frac{1}{2}\,\xi\,R\,\phi^2
-\frac{1}{4!}\,\lambda\,\phi^4
\nonumber\\
&-&\!\Lambda \m 
\frac12 M_p^2 \,R
+\alpha_{1} R_{\mu\nu\rho\sigma} R^{\mu\nu\rho\sigma}
 +\alpha_{2} \, R_{\mu\nu} R^{\mu\nu} +\alpha_{3}\,R^2\,\Big\}.
\eea

\medskip\noindent
 $M_p\!=\!(8\pi G)^{-1/2}$, 
$g\!\equiv\!\vert\det g_{\mu\nu}\vert$ and $\alpha_{1,2,3}$ and
$\xi$ are coupling  constants; we also set $\Lambda=0$.
We re-write (\ref{L}) in terms of the (square of) Weyl tensor ($C^2$) and Gauss-Bonnet term~($G$):
\smallskip
\bea\label{tensor}
R_{\mu\nu\rho\sigma}R^{\mu\nu\rho\sigma}&=& 2\,C^2-G+\frac{1}{3}\,R^2,
\nonumber\\
R_{\mu\nu} R^{\mu\nu} &=& \frac{1}{2} C^2 -\frac{1}{2}\,G +\frac{1}{3}\,R^2.
\eea

\medskip\noindent
Then 
\medskip
\bea\label{J0}
S= \int d^4 x \sqrt g\,\,\Big\{\,
\frac{1}{2} \, g^{\mu\nu} \partial_\mu\phi \partial_\nu \phi
\m\frac{1}{2} \,(M_p^2 \p \xi\,\phi^2) \,R -\frac{\lambda}{4!}\,\phi^4
+ \alpha\, R^2 
+\gamma\, C^2 
+\delta G\,\Big\},
 \eea  
where
\bea\label{bc}
\alpha=
\alpha_3+\frac{1}{3} (\alpha_1+\alpha_2)
,\qquad
 \gamma=  2\alpha_1+\frac{\alpha_2}{2}
,\qquad
\delta=  -\alpha_1-\frac{\alpha_2}{2}.
\eea

\medskip\noindent
The Gauss-Bonnet term  is  topological (total derivative);
the Weyl tensor is vanishing in the Friedmann-Robertson-Walker 
universe. Also, the coefficients $\gamma$, $\delta$ are
independent of $\xi$ and do not play a role in inflation.
Depending on the  values of the field $\phi$ and of 
$\xi$, $\alpha$, $\lambda$, eq.(\ref{J0}) covers both Higgs- and 
Starobinsky-inflation models which are  different limits of the 
same model as we briefly review  below (for a review see \cite{Riotto}).

One eliminates $R^2$ in (\ref{J0}) by replacing it with $R^2\ra \m 2\sigma^2 R-\sigma^4$
with $\sigma$ a new (auxiliary) scalar  field; its equation of motion $\sigma^2=\m R$
recovers (\ref{J0}) from a new, equivalent form
\medskip
\bea
S = \int d^4 x {\sqrt g}\,\,\Big\{\,
\frac{1}{2}\,g^{\mu\nu} \partial_\mu\phi\,\partial_\nu\phi
\m \frac{1}{2}\,M_p^2 \,f(\phi,\sigma)\,R
- W(\phi,\sigma)\,\Big\},
\eea

\medskip\noindent
where
\be\label{J}
 f(\phi,\sigma)=1+
\frac{1}{M_p^2}\Big[4\,\alpha\,\sigma^2 \p  \xi\,\phi^2  \Big],
\qquad
 W(\phi,\sigma)= \frac{1}{4!}\,\lambda\,\phi^4 + \alpha\,\sigma^4.
\ee

\medskip\noindent
Assume  $\alpha>0$, $\xi>0$, therefore $f>0$ and we rescale the  metric to\footnote{We use
$
\int d^4 x \sqrt g \,\,
(1/2)\, f\,R=\int d^4 x \sqrt {\hat g}\,\, (1/2)\,
\big[\hat R\,\m \, (3/2)\,  \hat g^{\mu\nu}\, 
\partial_\mu(\ln f)\,\,\partial_\nu(\ln f)
\big]$.
}
$\hat g_{\mu\nu}= f(\phi,\sigma) \,g_{\mu\nu}$.
Then the  action becomes (variables in the Einstein frame are marked with a hat):
\bea
\hat S\!=\!\!\int \!\! d^4 x \sqrt{\hat g}\,\,\Big\{
\frac{1}{2 f}\,
 \hat g^{\mu\nu} (\partial_\mu\phi)(\partial_\nu\phi)
-\frac{1}{ f^2}\, W(\phi,\sigma) 
\m
 \frac{M_p^2}{2} \hat R+\frac{3}{4} M_p^2 \,\hat g^{\mu\nu} \,\partial_\mu(\ln f)\,
\partial_\nu(\ln f)\Big\}.
\eea

\medskip\noindent
After a  field redefinition ($\sigma\ra \rho$) with $\rho$ a new real scalar field:
\bea\label{rho}
\ln f(\phi,\sigma)=q_0\,\rho, \quad q_0\equiv \frac{\sqrt{2/3}}{M_p},
\eea
then
\bea\label{E}
\hat S=\int d^4 x {\sqrt{\hat g}}\,\,
\Big\{\,\frac{1}{2}\,e^{-q_0\, \rho }
 \,\hat g^{\mu\nu} (\partial_\mu\phi)(\partial_\nu\phi)
- \hat W(\phi,\rho) 
\m \frac{1}{2}\,M_p^2\, \hat R
+\frac{1}{2}\, \hat g^{\mu\nu} \,(\partial_\mu\rho)(\partial_\nu\rho)\Big\},
\eea
with
\bea\label{EE}
\hat W(\phi,\rho)=
\frac{3}{4} M_p^2 \, M^2_\alpha  \Big[ 1- \Big(1 \p \frac{ \xi\,\phi^2}{M_p^2}\Big)
\,e^{-q_0\,\rho}\Big]^2
+
\frac{\lambda}{4!} \,\phi^4\,e^{- 2\,q_0 \,\rho}, 
\qquad M^2_\alpha=\frac{M_p^2}{12\,\alpha},
\eea
where scalaron $\rho$  enters as an exponent only. Eqs.(\ref{E}), (\ref{EE})
give the Einstein frame result.

One has Starobinsky inflation if $\lambda\!=\!\xi\!=\!0$ (Higgs field absent)
or $\xi\phi^2\!\ll\! M_p^2$ and $\alpha\!\sim\! 5\! \times\!  10^8$ \cite{AS3,Hawking3,Mass};
then  $\exp(q_0\rho)\!=\!1\!+\! 4\alpha\sigma^2/M_p^2$. For large $\sigma$
the first kinetic term vanishes and
\bea\label{si}
\hat W=\frac34 M_p^2\,M^2_\alpha\,  \big(1-e^{-q_0\,\rho}\big)^2, 
\qquad\qquad
 M^2_\alpha=\frac{M_p^2}{12\, \alpha}.
\eea

 One has Higgs inflation if
  $\alpha=0$ ($M_\alpha\!\ra\! \infty$) or $\alpha\,\sigma^2\!\ll \! M_p^2$ 
with $\xi\!\sim\! 1.8\!\times\!  10^4$ \cite{S}.
Then $\exp(q_0\rho)\!=\!1 \p\xi\phi^2/M_p^2$. For large $\phi$
the first kinetic term vanishes and the  potential is
\bea
\hat W=\frac34 M_p^2\,M^2_\xi \, \big(1-e^{-q_0\,\rho}\big)^2, 
\qquad\qquad
M^2_\xi=\frac{\lambda}{18 \,\xi^2} M_p^2.
\eea

Finally, consider the limit of large $\xi$ with $\alpha\not=0$.
In Jordan frame, the Higgs field kinetic term is subleading to  $\xi\,\phi^2 R$ and can be ignored
(at least during inflation). Then $\phi$ can be integrated
out via its equation of motion which is $\phi_c^2=-6\xi\,R/\lambda$. 
Then eq.(\ref{J0}) gives \cite{AS}
\medskip
\bea\label{eq14}
S=\int d^4 x \sqrt g\, \Big\{
\m \frac{1}{2} M_p^2\,R + \frac{M_p^2}{12\, M^2} \,R^2 
+\cdots\Big\},
\qquad M^2=\frac{M_p^2}{12 \,\big(\alpha +3\xi^2/(2\lambda)\big)}.
\eea
This is the ``usual'' Starobinsky model, which in the Einstein frame gives
eq.(\ref{si}) with $\alpha\!\ra\!\alpha\!+\!\frac{3 \xi^2}{(2\lambda)}$
and a  modified  scalaron mass $M\!\not=\!M_\alpha$.
The relation between $M$ and $M_\alpha$ is
\bea
M^2=\frac{M^2_\alpha}{1+ 18 \,(\xi^2/\lambda)\,M_\alpha^2/M_p^2}. 
\eea
To reproduce  the observed amplitude of curvature perturbations
$M\approx 1.3\times 10^{-5} M_p$ \cite{Mass}; this brings
a correlation among $\xi$, $\lambda$ and $\alpha$; if $\xi\!=\!\sqrt{\lambda/18}\,(M_p/M)$,
$M$ fixed, then $M_\alpha\!\ra\! \infty$.

\section{Starobinsky-Higgs model: quantum corrections}\label{quantum}

\subsection{Two-loop  effective action  with non-minimal coupling and $R^2$ terms}

Consider now the quantum corrections to action (\ref{L}); this is analytically continued 
to $d=4-2\epsilon$ dimensions, for regularization purposes.
The ``bare'' action  has a similar form. Note that the metric is not quantised.
The bare ($L_B$) and renormalized ($L$) Lagrangian
are related by $L_B\!=\mu^{2\epsilon} L$. 
At two-loop the counterterms $\delta L$ have the same form as (\ref{L}) 
($\!\delta\Lambda\!=\delta M_p=\!0$).

The quantum corrected potential $V$  is found by using its expansion 
in  powers of the curvature and its Callan-Symanzik (CS) equation:
\bea\label{c0}
\Big(\mu\frac{\partial}{\partial \mu}
+\beta_\lambda\,\frac{\partial}{\partial\lambda}+\beta_\xi \frac{\partial}{\partial \xi}
-\gamma\,\phi\, \frac{\partial}{\partial\phi}
+\sum_{j=1}^3 \beta_{\alpha_j}\frac{\partial}{\partial \alpha_j}\Big)\, V(\phi)=0,
\eea

\medskip\noindent
where $\beta_X$ is the beta function of the coupling $X$ and $\gamma$ is the anomalous 
dimension of $\phi$.

The  two-loop  solution to (\ref{c0}) with terms quadratic in
curvature is found  by using the  method of \cite{O1,PT}.
Then
\medskip
\bea
V(\phi)=V_0(\phi)+V_1(\phi)+V_2(\phi).
\eea

\medskip\noindent
$V_0$ is the potential in flat space-time,   $V_1$ is linear in $R$,
  $V_2$ contain terms quadratic $R$, $R_{\mu\nu}$ and  $R_{\mu\nu\rho\sigma}$.
We ignore higher powers of $R$, etc.
 From  eq.(\ref{c0}) by grouping terms of similar structure  (linear in $R$, etc)
one  finds three  independent CS equations  for $V_0$, $V_1$, $V_2$. These are solved
iteratively  at two-loop (order by order), see Appendix~A for   details. One has
\medskip
\be\label{op}
V_0=\!\frac{\tilde\lambda(\phi)}{4!}\phi^4,\quad
V_1 =\! \frac{\tilde \xi(\phi)}{2} R\,\phi^2,\quad
V_2= -\tilde\alpha_1(\phi)  R_{\mu\nu\rho\sigma} R^{\mu\nu\rho\sigma}\!
-\tilde\alpha_2(\phi) R_{\mu\nu} R^{\mu\nu}\!
-\tilde\alpha_3(\phi) R^2.\,\,
\ee

\medskip\noindent
 $\tilde\lambda$, $\tilde\xi$ and $\tilde\alpha_{1,2,3}$ are functions\footnote{
To simplify notation, below we do not  write explicitly their argument $\phi$.} of $\phi$:
\bea
\tilde\lambda=\tilde\lambda(\phi),
\qquad
\tilde\xi=\tilde\xi(\phi),
\qquad
\tilde\alpha_i=\tilde\alpha_i(\phi).
\eea
At two-loop level, from (\ref{vflat}) 
\medskip
\bea\label{tildelambda}
\qquad\tilde\lambda=\lambda \Big\{1+ 
\frac{3\lambda}{2\kappa} \,\Big[\ln \frac{\phi^2}{\mu^2}-\frac{25}{6}\,\Big]
+
\frac{3\lambda^2}{4\kappa^2}\,\Big[\,\frac{515}{6} -29\ln \frac{\phi^2}{\mu^2} 
+3\ln^2 \frac{\phi^2}{\mu^2}\,\Big]\Big\}, 
\eea
%
with $\kappa=(4\pi)^2$. From (\ref{vone})
\bea\label{tildexi}
\frac{\tilde\xi}{2}\!\!\! & = &\!\!\!
 \Big\{\,
\frac{\xi}{2} + \frac{\lambda}{4\,\kappa} \,\Big(\xi \p \frac{1}{6}\Big)
\Big(\ln\frac{\phi^2}{\mu^2}-3\Big) 
-\frac{\lambda^2}{4\,\kappa^2}\,\Big(\frac{7}{6}\,\xi \p \frac{13}{36}\Big)
 \,\Big(\ln\frac{\phi^2}{\mu^2}-3\Big) 
\nonumber\\
&+&\!\!\!\!
\frac{2\lambda^2}{\kappa^2} \frac{1}{4} \Big(\xi  \p \frac{1}{6}\Big) 
\Big(7-3\ln\frac{\phi^2}{\mu^2} +\frac{1}{2}\ln^2 \frac{\phi^2}{\mu^2} \Big),
\Big\}
\eea
and finally, from eqs.(\ref{vtwo}), (\ref{vtwop})
\bea\label{in}
&& -\tilde\alpha_j=  
-\alpha_j+ (-1)^{j+1} \frac{1}{180\kappa}\,\ln\frac{\phi}{\mu}, \quad j=1,2;\qquad
\nonumber\\
&&-\tilde\alpha_3= 
-\alpha_3+\frac{1}{2\kappa} \,\Big(\xi \p \frac{1}{6}\Big)^2 \ln\frac{\phi}{\mu}
+  \frac{\lambda}{2 \kappa^2} 
   \Big(\xi \p \frac{1}{6}\Big)^2 \ln^2\frac{\phi}{\mu}.
\eea

\medskip\noindent
To solve for $V_0$, $V_1$, $V_2$,  the following
 boundary conditions were used
\medskip
\bea
V_0^{(4)}\Big\vert_{t=0}=\lambda,\qquad 
V_1^{(2)}\Big\vert_{t=0}= R\, \xi,\qquad
\tilde\alpha_j\Big\vert_{t=0}  
=\alpha_j, \, (j:1,2,3),\quad t\equiv \ln\frac{\mu}{\phi}.
\eea

\medskip\noindent
Here  $V_j^{(k)}$ is the $k$-th derivative of $V_j$ with respect to $\phi$. 
Note that $\tilde\xi$, $\tilde\lambda$, $\tilde\alpha_j$  are all functions of $t$
and that  $V$  simplifies  in the  one-loop conformal limit of $\xi=\m 1/6$.

With the above expressions the action becomes
\medskip
\bea\label{two}
S&=&\int d^d x {\sqrt g}\,\,\Big\{\,\,
\frac{1}{2}\,Z_\phi g^{\mu\nu} \partial_\mu\phi\partial_\nu\phi
\m \frac{1}{2} \,\big(M_p^2\p \tilde\xi\,\phi^2\big)\,R -\frac{\tilde\lambda}{4!}\,\phi^4
\nonumber\\
&& \qquad\qquad+\,\,  
\tilde\alpha_1\,R_{\mu\nu\rho\sigma}\,R^{\mu\nu\rho\sigma}
+\tilde\alpha_2 \,R_{\mu\nu} R^{\mu\nu}
+\tilde\alpha_3\,R^2 \,\Big\}.
\eea

\medskip\noindent
Using eq.(\ref{tensor}), the $R^2$-term  acquires an extra  correction
\medskip
\bea\label{J1}
S= \int d^d x \sqrt g\,\,\Big\{\,
\frac{1}{2}\,Z_\phi \,(\partial_\mu\phi)^2
\m  \frac{1}{2} \,\big(M_p^2\p \tilde\xi\,\phi^2\big)\,R -\frac{\tilde\lambda}{4!}\,\phi^4
+\tilde \alpha\,R^2 
+\tilde\gamma\, C^2 
+ \tilde\delta G\,\Big\},
 \eea  
where
\bea\label{bcp}
\tilde \alpha=
\tilde\alpha_3+\frac{1}{3} (\alpha_1+\alpha_2)
,\qquad
\tilde \gamma=  2\tilde\alpha_1+\frac{\tilde\alpha_2}{2}
,\qquad
\tilde \delta=  -\tilde\alpha_1-\frac{\tilde\alpha_2}{2}
\eea
The action in (\ref{J1}), (\ref{bcp}) is a combination of the
Higgs and Starobinsky inflationary models. Here $\tilde\xi$, $\tilde\alpha$ 
and $\tilde \lambda$ are field-dependent,  two-loop corrected  couplings,
also  $\tilde\gamma$,  $\tilde \delta$ acquired $\phi$-dependence but
remain $\xi$-independent; their $\phi$-dependence may still play  a role e.g. via 
eqs of motion, if one integrates $\phi$.

Consider now the limit of (\ref{J1}) when the $R^2$-term has 
a vanishing classical coupling $\alpha=0$, see (\ref{bc}).
With a subtraction scale $\mu\!\sim\! M_p$ and  $\phi\ll\! M_p$,
 then $\tilde\alpha(\phi)\!>\!0$ and
\bea\label{qq}
\tilde\alpha(\phi) \,
=\frac{1}{2\kappa}\,\Big(\xi\p \frac16\Big)^2\ln\frac{M_p}{\phi}
+\text{two-loop}.
\eea

\medskip\noindent
For  $\phi$ fixed, small enough 
$\phi^2_0\! \ll\! M_p^2/\tilde\xi$, action (\ref{J1}) is dominated by: 
 $(1/2)\, M_p^2 R+\tilde \alpha(\phi_0)\,R^2$.

Therefore, we find an interesting result: even if it is absent at the classical level,
the ``usual'' Starobinsky inflation is generated at the {\it quantum level}, if
 there exists  a suitable (large) non-minimal coupling $\xi$ of the Higgs field
(which otherwise plays no role in inflation), see also \cite{C2}.
This requires tuning   $\xi$ (rather than $\alpha$ \cite{Hawking3}), 
with $\tilde\xi\sim\xi$ for small $\lambda$.

\subsection{Large $\xi$ limit: scalaron quantum action after integrating $\phi$}
\label{largexi}

Let us consider the limit of large $\xi$ in  the quantum action eq.(\ref{J1}),
integrate $\phi$ analytically and then  find the effective action. 
 This is similar  to the classical discussion near eq.(\ref{eq14});
at large $\xi$ the kinetic term of $\phi$  is  subleading to similar 
two-derivative terms coming  from $\xi\phi^2\,R$  after integration by parts
and can be ignored. Then $\phi$ is integrated via its equation of motion
 $\phi^2_c=6\xi\,\vert R\vert/\lambda$. One can still use $\phi_c$ at the one-loop level,
 as we  do below.

After integrating $\phi$, the action in (\ref{J1}) becomes 
(ignoring hereafter the  $C^2$ and $G$-terms)
\bea\label{ss}
S=\int d^4 x \,\sqrt g\,\Big\{
\,
 \frac{1}{2}\,M_p^2\,F(\m R)
\Big\},
\eea
where
\bea\label{fofR}
F(R)= R +\frac{2\,R^2}{M_p^2}\,
\gamma_0 \,\Big[1 + \gamma_1 + \gamma_2\,\ln \frac{6\xi \vert R\vert}{\lambda\mu^2}
+\gamma_3\,\ln^2\frac{6\xi \vert R\vert}{\lambda\mu^2}\,\Big],
\eea

\medskip\noindent
and where $\gamma_{0,1,2,3}$ are read from (\ref{tildelambda}), 
(\ref{tildexi}), (\ref{in}). At one-loop
\bea
\gamma_0=\alpha+3\xi^2/(2\lambda),\quad
\gamma_1=\frac{3\,\xi\,(13\,\xi \m 2)}{8\,\kappa\,\gamma_0},
\quad
\gamma_2= -\frac{1}{\kappa\,\gamma_0} (\xi^2-\xi/6+1/144),
\quad 
\gamma_3=0.
\eea

\medskip
 Note the $R^2\ln\vert R\vert$  term generated  by integrating matter field $\phi$;
 $R^2\ln^2\vert R\vert$ terms are also generated at two-loops. 
Integrating $\phi$ in (\ref{J1}) also induces a ``mixing'', ignored below,
 $(-C^2+G/3)\ln (\xi\vert R\vert/\mu^2)$ since  $\tilde\gamma$, 
$\tilde\delta$ depend on $\phi$ (see also \cite{ZL}).
From (\ref{fofR}) for $\xi^2\!\gg\! \lambda\alpha$ 
\medskip
\bea\label{qs}
S=\int d^4x \sqrt g\,  \Big\{
\m \frac12\, M_p^2 R+ \frac{M_p^2}{12 \,M^2}\,R^2\Big[1+\frac{13\lambda}{4\kappa}
-\frac{2 \lambda}{3\kappa}\,\ln\frac{(6\xi \vert R\vert)}{\lambda\,\mu^2}\Big]
+\cO(\alpha\lambda/\xi^2)\Big\},
\eea

\medskip\noindent  
with $M$ as in (\ref{eq14}).
Eqs.(\ref{ss}) to  (\ref{qs}) give
a one-loop corrected version of the  classical action  in (\ref{eq14}).
A similar action was analysed recently for cosmological consequences\footnote{
For a similar action, of different, quantum gravitational   origin, see \cite{QG}.}
 in \cite{AS2,Lee,My1,My2,My3}, 
\bea
S=\int d^4 \sqrt{g}\,\Big\{ 
\m \frac12\, M_p^2 R+ \frac{a}{2} R^2\,\frac{1}{
\big[ 1 + b \ln (\vert R\vert/\mu^2)\big]}\Big\},
\eea
and it  was found that  successful inflation demands 
$a\sim 10^8-10^9$ and $b\leq 10^{-2}$ 
(for the spectral index to agree with observations) \cite{AS2}. In our case
$a=2\gamma_0 [1+\gamma_1+\ln (6\xi/\lambda)]$ and $b=-\gamma_2/(1+\ln (6\xi/\lambda))$.
If $\xi^2\gg \lambda\alpha$, then
 $a=3\xi^2/\lambda$ and $b=2\lambda/(3\kappa)\approx 4.2\times 10^{-3}\lambda$.
The constraint for $a$ is respected by suitable $\xi \sim 10^4\sqrt\lambda$, while that 
of $b$ is easily satisfied for perturbative $\lambda$. 
Increasing $b$, which is of quantum origin here, leads to an increase
of the scalar-to-tensor ratio \cite{AS2}.  
From  $b$ one also can infer constraints on the matter content. 

To find the Einstein frame potential,
eqs.(\ref{ss}), (\ref{qs}) can be written as \cite{SM}
\bea
\label{s2}
S=
\int d^4 x \sqrt g\,
\,\frac{1}{2}\,M_P^2\,\Big[\m  F'(\sigma) R +F(\sigma)-\sigma F'(\sigma)\Big],
\eea

\medskip\noindent
The equation of motion for auxiliary field $\sigma$ is $\sigma\!=\!\m R >0$ and if used in (\ref{s2}) 
recovers (\ref{ss}). Further, in (\ref{s2}) 
rescale the metric followed by a field redefinition $\sigma\ra\rho$, using 
$\hat g_{\mu\nu}=F'(\sigma) g_{\mu\nu}$ and 
$\ln F'(\sigma)=q_0\,\rho$ where, as usual $q_0\equiv \sqrt{(2/3)} (1/M_p)$; also $F^\prime(\sigma)>0$.
We then find  the action for $\rho$ in the Einstein frame
\medskip
\bea
\hat S=\int d^4x \sqrt{\hat g}\,\,\Big\{\,\m  \frac12 M_p^2\,\hat R
+\frac 12 \,\hat g^{\mu\nu}\,\partial_\mu\rho\,\partial_\nu\rho -\hat W_\text{eff}(\rho)\Big\},\quad
\eea
with
\bea\label{ww}
\hat W_{\text{eff}}(\rho)=\frac12\,M_p^2\,\frac{\sigma}{F'(\sigma)}\,\Big\{1-\frac{1}{\sigma}
\frac{F(\sigma)}{F'(\sigma)}\Big\}_{\sigma=\m R(\rho)}
\eea

\medskip\noindent
As shown in (\ref{ww}),  $\sigma$ is replaced by the solution of the equation: 
$\ln F'(\m R)=q_0\rho$.
To find the solution $R(\rho)$  we search for one of the form
\medskip
\bea\label{ee}
R=R_0\,\Big[1+
\delta_1  + 
\delta_2\,\ln \frac{6\xi \vert R_0\vert }{\lambda \mu^2}\,\Big],\quad
R_0\equiv \m 
\frac{M_p^2}{4 \,\gamma_{0}}\,(e^{q_0\,\rho}-1).
\eea

\medskip\noindent
$R_0$  is the solution at the tree-level, 
in which $\cO(1/\kappa^n)$ terms, $n\geq 1$ are ignored. Using (\ref{fofR})   we find
$\delta_1=-(\gamma_1+\gamma_2/2)$, 
$\delta_2=-\gamma_2$.
With  (\ref{ee}) the scalaron effective potential is 
\medskip
\bea
\hat W_{\text{eff}}(\rho)=\frac{3}{4}\,M_p^2\,M^2\,\Big[ 1-e^{-q_0\,\rho}\Big]^2
\Big\{ 1
- \gamma_1 
- \gamma_2  
\Big[ 
\ln\Big( \frac{3 \,\xi \,M_p^2 }{2\lambda\, \gamma_0  
 \,\mu^2}\Big)
+q_0\rho+\ln \big[1- e^{- q_0\rho} \big]\Big]\Big\},
\eea

\bigskip\noindent
with $M$ defined in eq.(\ref{eq14}). Notice  the correction  linear in $\rho$. The potential contains
a dependence on the subtraction scale $\mu$, as expected. 
Further, if $\xi^2\gg \lambda\,\alpha$ then
\medskip
\bea\label{eqf}
\hat W_\text{eff}(\rho)
=
\frac{3}{4} M_p^2 \,M_\mu^2 \Big[1-e^{-q_0\rho}\Big]^2 
\Big\{1-\frac{13\lambda}{4\kappa} 
+\frac{2\lambda}{3\kappa}\,q_0\,\rho
+\frac{2\lambda}{3\kappa}\ln \big[ 1-e^{-q_0\rho}\big]
\Big\},
\eea
with 
\bea
M_\mu^2=M^2 \Big[1+\frac{2\lambda}{3\kappa}\ln\frac{M_p^2/\xi}{\mu^2}\Big].
\eea

\medskip\noindent
 Eq.(\ref{eqf}) is the Einstein frame counterpart to eq.(\ref{qs}).
The dependence of $\hat W_\text{eff}$ on the  scale $\mu$ was included in
 $M_\mu$; this  may be regarded as the  ``running''  scalaron mass, with an
``emergent''  UV cutoff  $M_p/\sqrt\xi$ in the Einstein frame.
The new term  linear in scalaron ($\propto\rho$)
is important near/above the Planck scale  $M_p$ but its effect is 
reduced by a small $\lambda$ (of the Standard Model).  These corrections correspond  to the 
$R^2\ln (\xi \vert R\vert/\mu^2)$ term in  the Jordan frame. 
In the absence of these terms (e.g. if $\lambda\ra 0$) we  recover
 the ``standard''  Starobinsky inflation. This analysis can be repeated 
in  the two-loop action,  using a one-loop solution of the  equation of 
motion of $\phi$.

\subsection{Einstein frame action}

The general two-loop result in eq.(\ref{J1}) can be mapped to the Einstein frame.  
We eliminate the $R^2$ dependence  via a  replacement 
$R^2\ra\m 2\sigma^2\,R-\sigma^4$ in (\ref{bcp}) where $\sigma$ is a new real scalar field. 
Its equation of motion $\sigma^2=\m R$ recovers (\ref{J1})  from a new, equivalent 
action\footnote{We ignore here the additional  terms
 $\tilde\gamma\, C^2+\tilde\delta\, G$ in (\ref{J1}).}:
\medskip
\bea
S = \int d^d x {\sqrt g}\,\,\Big\{\,
\frac{1}{2}\,Z_\phi \,(\partial_\mu\phi)^2
\m  \frac{1}{2}\,M_p^2 \,\tilde f(\phi,\sigma)\,R
-\tilde W(\phi,\sigma)\,\Big\},
\eea

\medskip\noindent
where
\be\label{JE}
\tilde f(\phi,\sigma)=1+
\frac{1}{M_p^2}\Big[4\,\tilde\alpha\,\sigma^2\p  \tilde\xi\,\phi^2  \Big],
\qquad
\tilde W(\phi,\sigma)= \frac{1}{4!}\,\tilde\lambda\,\phi^4 + \tilde\alpha\,
\sigma^4.
\ee

\medskip\noindent
Avoiding ghosts requires  $\tilde f>0$, which is true if $\xi, \alpha>0$ as assumed
(the determinant of the matrix in field space
 $\phi$, $\sigma$  is then positive). We follow previous  steps and 
rescale the metric to  $\hat g_{\mu\nu}=\tilde f(\phi,\sigma) \,g_{\mu\nu}$,
then redefine $\ln \tilde f(\phi,\sigma)\!=\! q_0\,\rho$ with   $q_0\!\equiv\!\sqrt{(2/3)}/M_p$.
Then
\medskip
\bea\label{f1}
\hat S=\int d^dx {\sqrt{\hat g}}\,\,
\Big\{\,\frac{1}{2}\,Z_\phi\,e^{-q_0\, \rho }
 \,\hat g^{\mu\nu} (\partial_\mu\phi)(\partial_\nu\phi)
- \hat W(\phi,\rho) 
\m 
\frac{1}{2}\,M_p^2\, \hat R
+\frac{1}{2}\, \hat g^{\mu\nu} \,(\partial_\mu\rho)(\partial_\nu\rho)\Big\},
\eea
with
\bea\label{f2}
\hat W =  
\frac{3}{4} M_p^2 \,\tilde M(\phi)^2  
\Big[ 1- \Big(1\p  \frac{\tilde \xi(\phi)\,\phi^2}{M_p^2}\Big)
\,e^{-q_0\,\rho}\Big]^2
\!\!+\frac{\tilde\lambda(\phi)}{4!} \,\phi^4\,e^{- 2\,q_0 \,\rho},
\quad
\tilde M^2(\phi)\!=\!\frac{M_p^2}{12\,\tilde\alpha(\phi)},
\eea

\medskip\noindent
in the  Einstein frame. 
We made explicit the $\phi$-dependence of $\tilde M$, $\tilde\xi$, $\tilde\lambda$,
$\tilde \alpha$. As at tree-level,  $\rho$  shows up  only as an exponent 
that suppresses the kinetic term of $\phi$. 

Eqs.(\ref{f1}), (\ref{f2}) are the counterpart to the Jordan frame 
quantum result of eq.(\ref{J1}), in terms of $\tilde M$, $\tilde\xi$, 
$\tilde\lambda$, see also the classical  version eqs.(\ref{E}), (\ref{EE}).
 Either the Higgs field or scalaron or their mixing plays the 
inflaton role, depending on the relative size of these couplings at quantum  level.
We thus have a standard two-field inflationary model, see e.g. \cite{Lalak},
 with diagonal  metric $\gamma_{ab}$ in the field space ($\phi,\rho$) but
 with $\gamma_{\phi\phi}=e^{-q_0\rho}$ and $\gamma_{\rho\rho}=1$.

Finally, consider a  special case of  $\alpha=0$ i.e. no classical $R^2$-term.
 With  $\phi$ fixed $\phi=\phi_0$ and small enough 
 $\tilde\xi\phi_0^2\ll M_p^2$, then from (\ref{f2})
\medskip
\bea\label{pp}
\hat W=\frac{3}{4} M_p^2\,\tilde M(\phi_0)^2\Big[1-e^{-q_0\,\rho}\Big]^2,
\qquad 
\tilde M(\phi_0)^2\approx \frac{M_p^2}{12\,(\xi^2/\kappa)\,\ln (\mu^2/\phi_0^2)}.
\eea

\medskip\noindent
With subtraction scale $\mu$ chosen of order $M_P$ and $\phi_0\ll \mu$,
eq.(\ref{pp}) gives a quantum origin to $\tilde M(\phi_0)$ and to 
Starobinsky inflation if $(\xi^2/\kappa)\ln\mu^2/\phi_0^2\!\sim\! 5\times 10^8$
or  $\xi\!\approx\! 3.5\times 10^4$ ($\phi_0\!\sim\! m_Z$),
(which is also  that needed for Higgs inflation).
For this to happen, the  non-minimal coupling $\xi$ is essential. 
This  is the Einstein-frame counterpart to  the Jordan frame result, 
eq.(\ref{qq}).

\subsection{The behaviour of the quantum potential}\label{pheno}

Let us  examine  the potential in  eqs.(\ref{f2})
and compare it  against its classical behaviour  \cite{AS}.
The Hubble parameter is\footnote{In general 
$H^2=1/(3 M_p^2)\big(\frac12 \gamma_{ab}\dot\phi^a \dot\phi^b+W\big)$,
$\epsilon=\dot\phi_0^2/(2 M_p^2\,H^2)$,
$\phi_0^2\equiv \gamma_{ab}\dot\phi^a\dot\phi^b$.}
\bea\label{H2}
H^2=\frac{1}{3\,M_p^2}\Big(\frac12 \,\dot\phi^2\,e^{-q_0\,\rho}+\frac12 \dot\rho^2+\hat W\Big),
\eea
and 
\bea
\dot H=-\frac{1}{2\,M_p^2}\,\Big(\dot\phi^2 e^{-q_0\rho} +\dot\rho^2\Big).
\eea

\medskip\noindent
One has slow-roll conditions $\epsilon\!\ll\! 1$, $\eta_{\parallel}\!\ll\! 1$ and
spectral index $n_s=1+2\eta_{\parallel}-4\epsilon$ \cite{AS}, where
\medskip
\be\label{param}
\epsilon \equiv
-\frac{\dot H}{H^2}
=\frac{\big( \dot\phi^2\,e^{-q_0\,\rho}\!+\dot\rho^2\big)}{2\,M_p^2 \,H^2},
\qquad
\eta_{\parallel}\equiv \frac{-\ddot\phi_0}{H\,\dot\phi_0},
\qquad \text{with}
\qquad
\dot\phi_0^2\equiv 
\gamma_{ab}\dot\phi^a\dot\phi^b,\,\, (a, b\!:\phi,\rho).\,\,
\ee

\medskip\noindent
For slow roll regime and with $\hat W$ of (\ref{f2}), 
one  has $H^2\approx 1/(3M_p^2) \,\hat W$.
Then for larger values of the field $\rho$  and large  
$\tilde\xi(\phi)\phi^2> M_p^2$   
the bracket\footnote{For the other limit, of large $\rho$ and small $\tilde\xi$, 
 $\tilde W\approx (3/4) M_p^2\tilde M(\phi)^2\approx (3/4) M_p^2/(12 \,\alpha)\approx$ constant.}
in $\hat W$ that multiplies  $e^{-q_0\rho}$ is approximated by the term $\propto \tilde\xi$. 
To compensate this  effect 
and to keep $H$ constant (so that the  amplitude of curvature perturbations is unchanged) one 
has to  increase $\tilde M(\phi)$. 
 This is possible by considering a smaller  $\tilde\alpha(\phi)$ in 
 eq.(\ref{f2}). Therefore  $\tilde\xi$ and $\tilde\alpha\sim \alpha$ play
somewhat opposite roles while keeping $H$ unchanged, and they encode 
the effects of $\phi$ and $\rho$,  respectively; 
$\tilde\xi$ and $\tilde\alpha$ are largely  controlled by their tree-level values, 
$\xi$ and $\alpha$, and to a smaller extent by $\phi$,  $\lambda$ and the
renormalization scale $\mu$ entering in the  quantum corrections 
of eqs.(\ref{tildelambda}), (\ref{tildexi}), (\ref{in}).

\begin{figure}
\begin{center}
\includegraphics[width=6.cm,height=5cm]{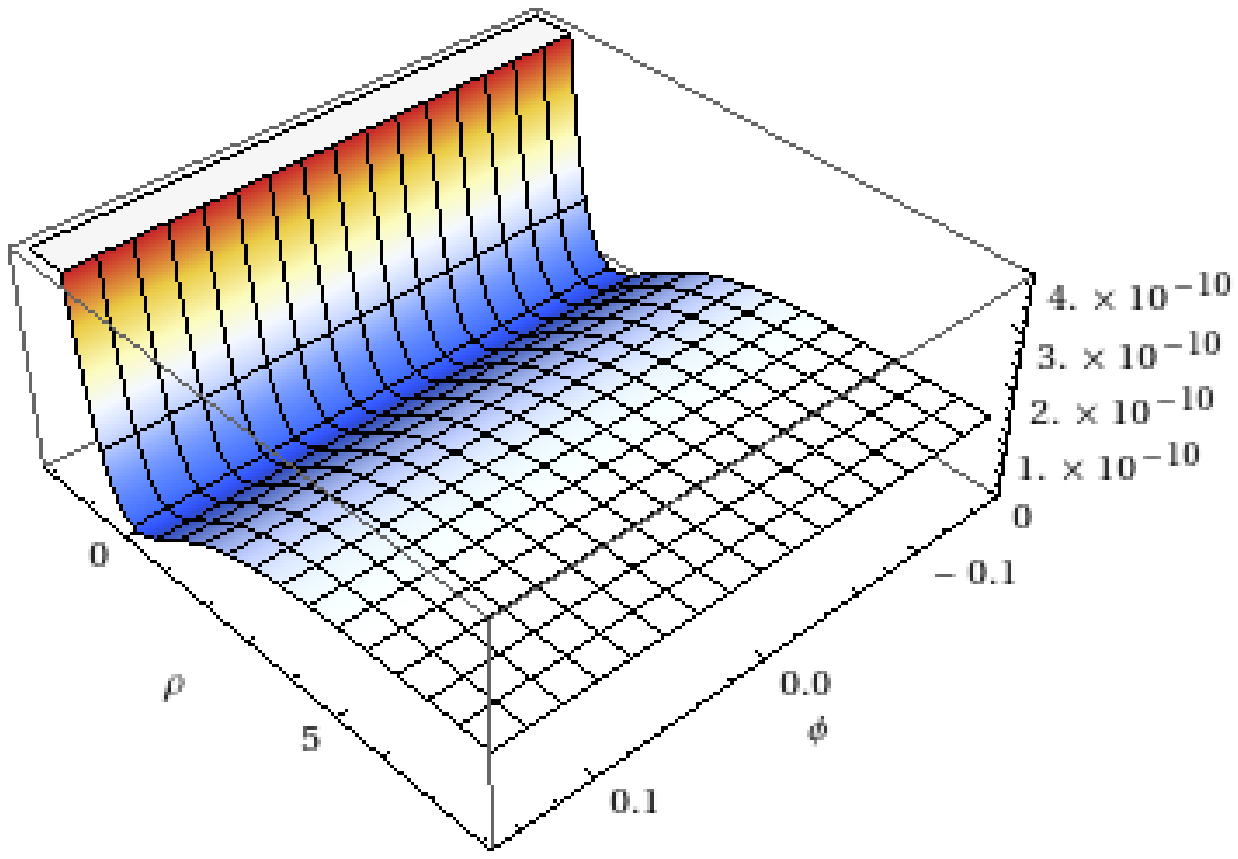}
\includegraphics[width=6.cm,height=5cm]{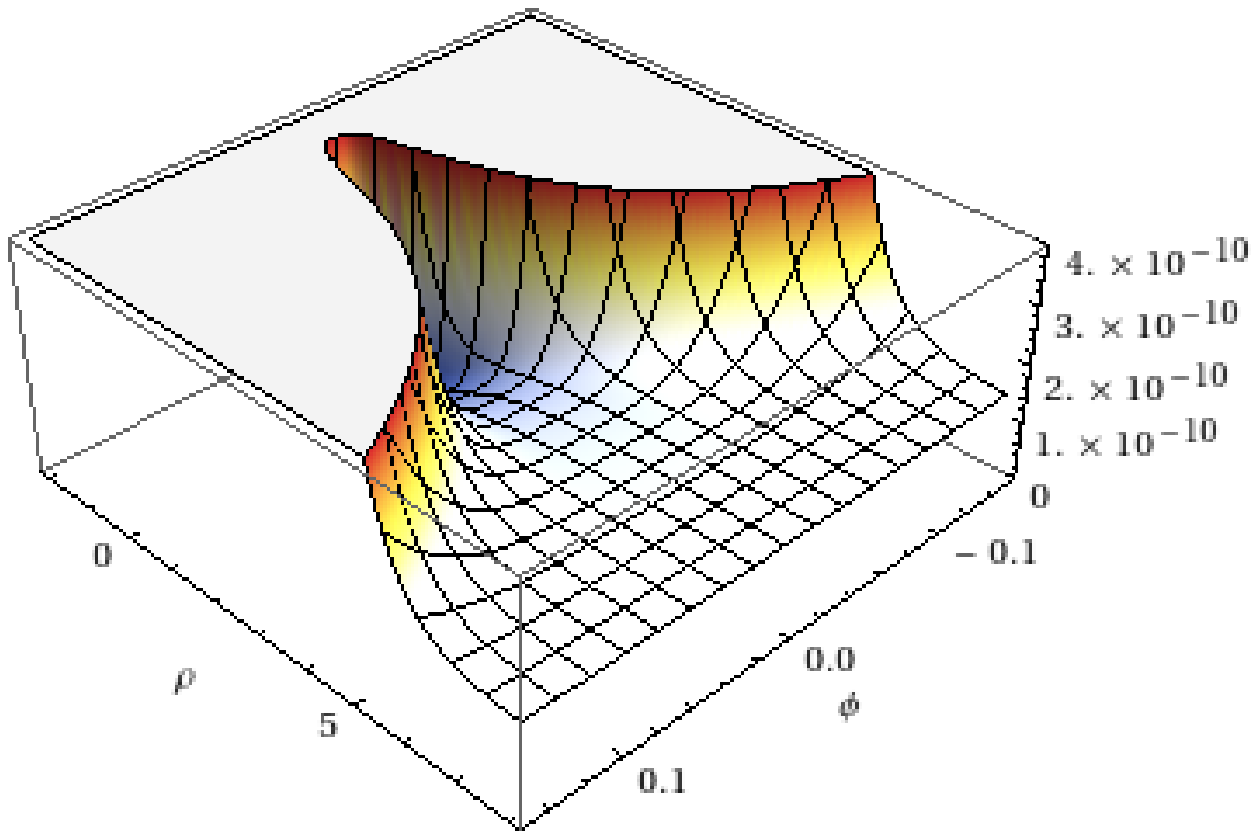}
\\
{\footnotesize{\hspace{0.7cm}  (a1)  \hspace{5.cm} (a2)}}
\\
\includegraphics[width=6.cm,height=5cm]{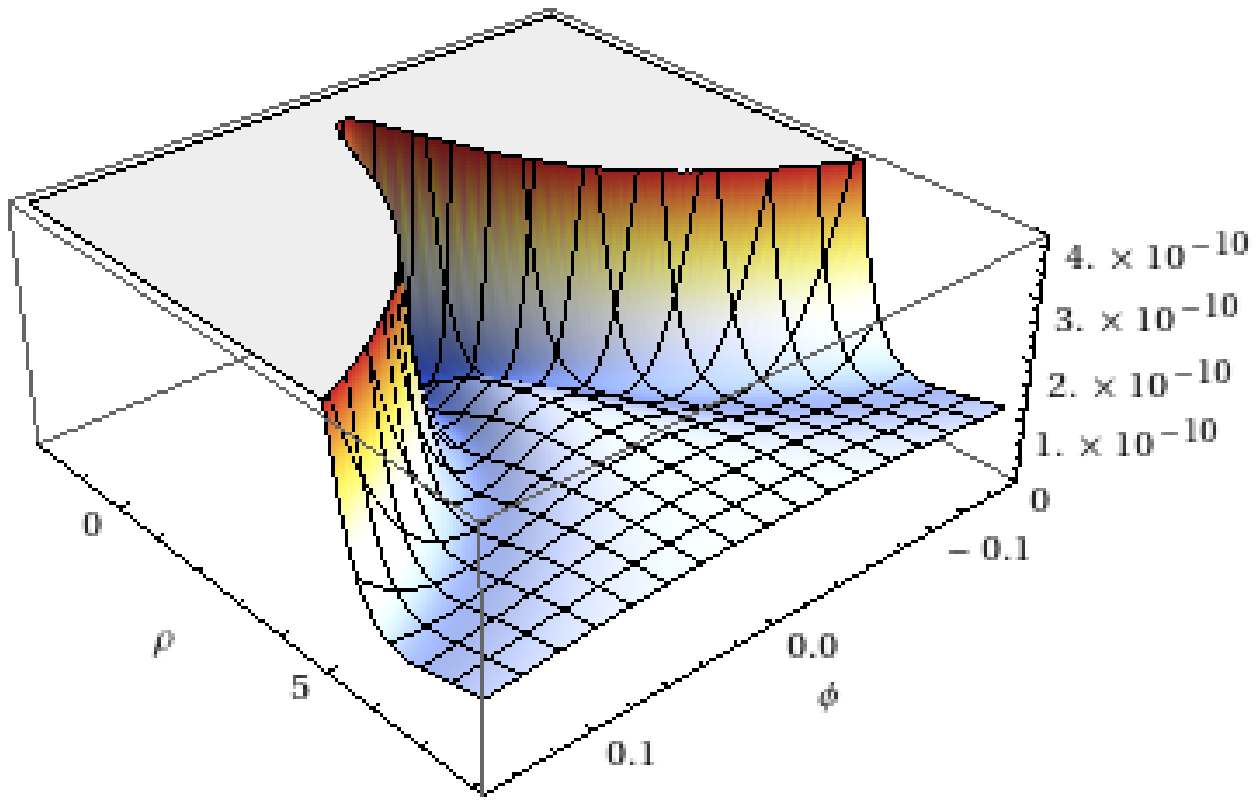}
\includegraphics[width=6.cm,height=5cm]{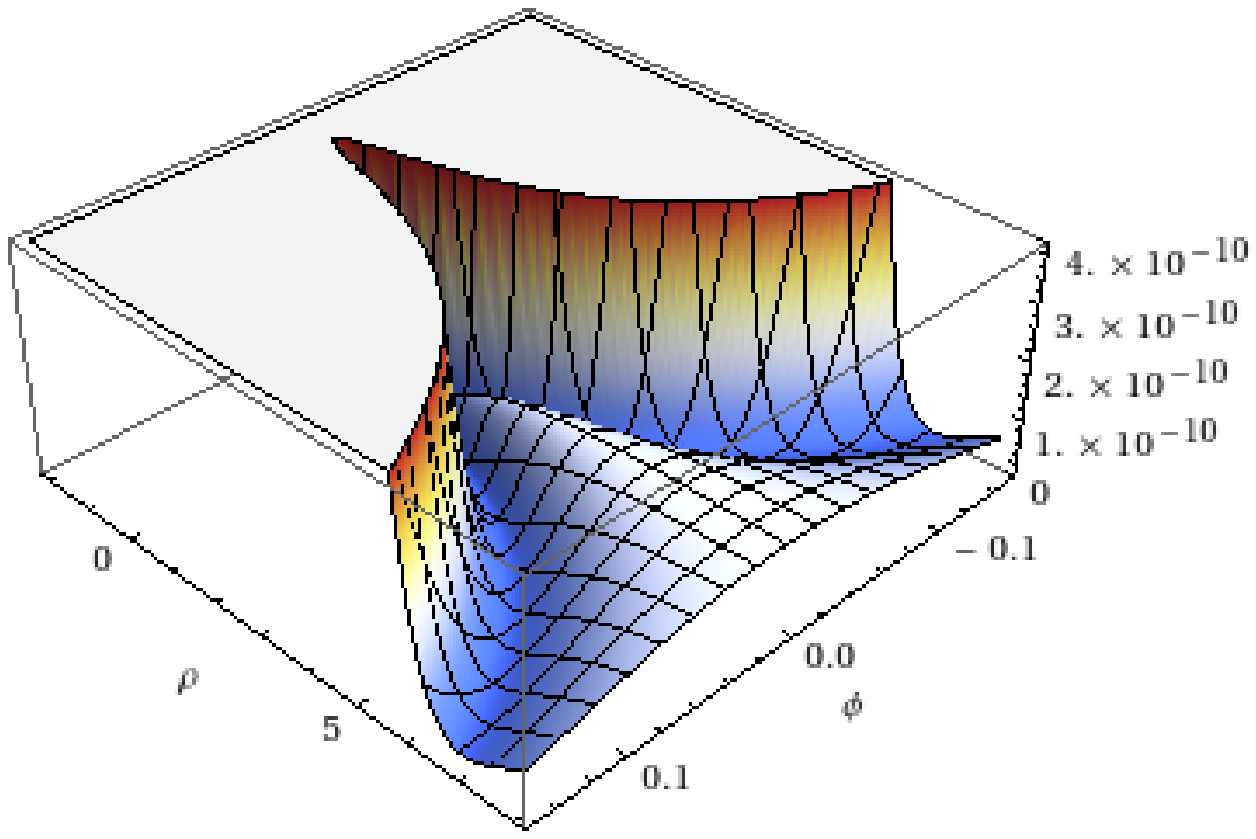}
\\
{\footnotesize{\hspace{0.7cm} (b1)  \hspace{5.cm} (b2)}}
\\
\includegraphics[width=6.cm,height=5cm]{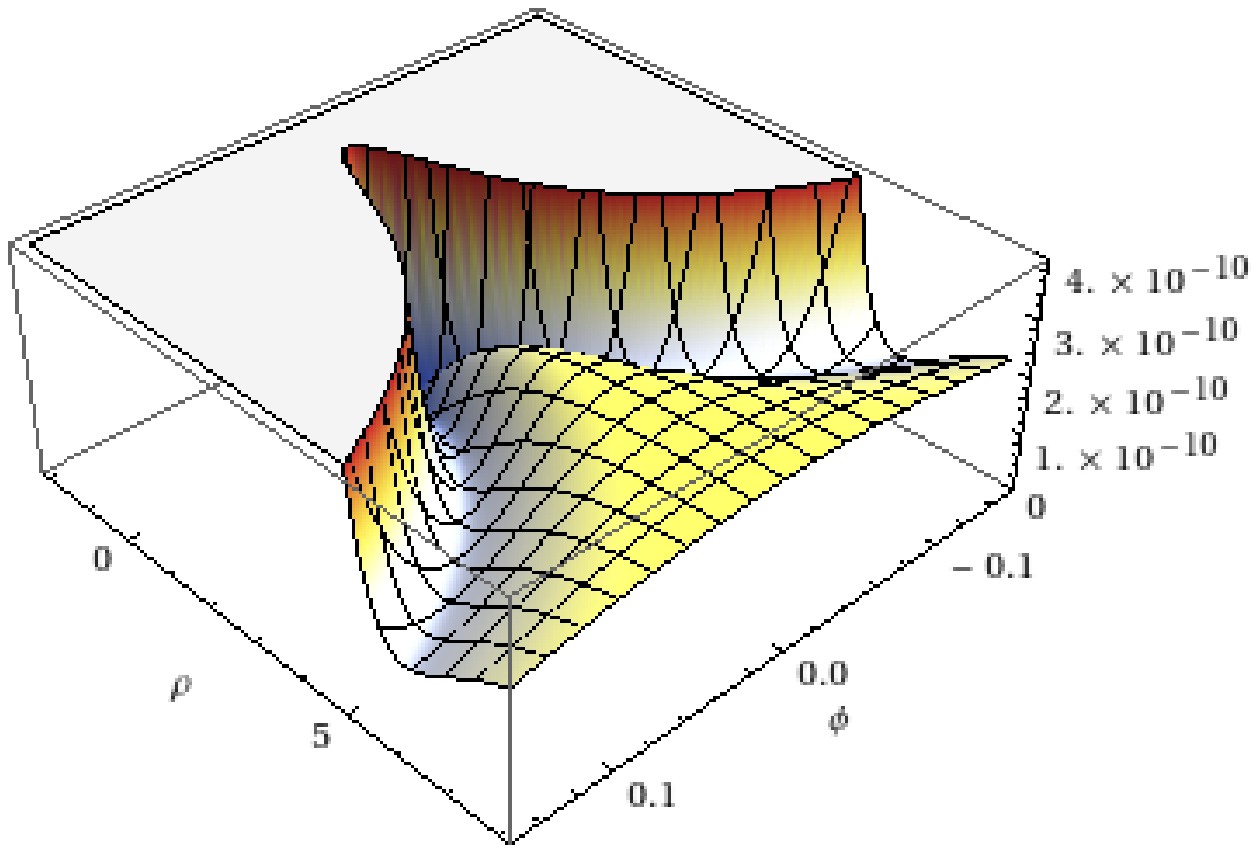}
\includegraphics[width=6.cm,height=5cm]{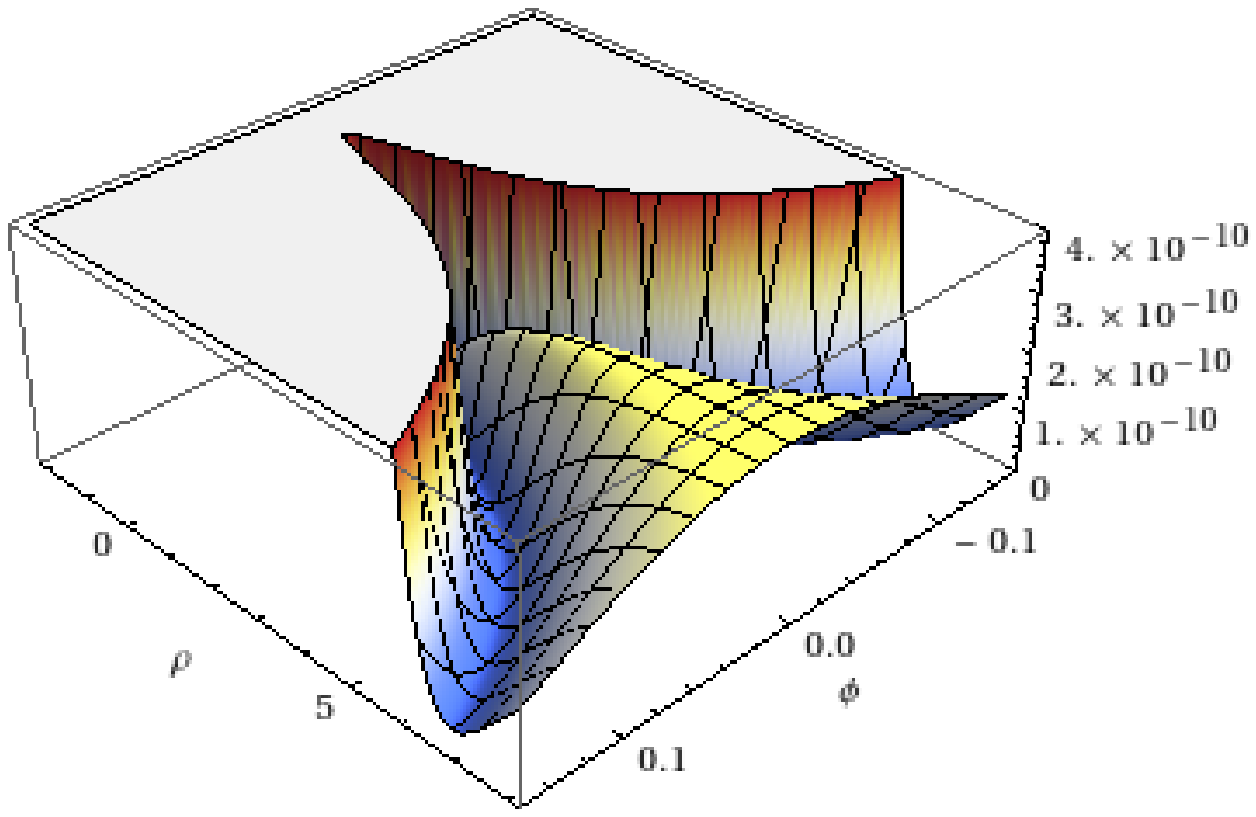}
\\
{\footnotesize{\hspace{0.7cm} (c1)  \hspace{5.cm} (c2)}}
\end{center}
\renewcommand{\baselinestretch}{0.9}
\vspace{-0.2cm}
\caption{\small 
The potential $\hat W(\phi,\rho)$  eq.(\ref{f2}) with  
values in Planck units and  subtraction scale $\mu=M_p$.
Plot {\bf (a1):} $\alpha\!=\!4.2\!\times\! 10^8$, $\xi=0$ and
$\lambda=0$ (the ``usual'' Starobinsky model).
Next consider $\lambda=0.06$ and:
{\bf (a2):} $\alpha=4.2\times 10^8$,  $\xi=0$,  so increasing $\lambda$ 
lifts up the valley at $\rho=0$.\,\,\, 
{\bf (b1):} $\alpha=4.2\times 10^8$, $\xi= 3000$,\,\,
{\bf (b2):} $\alpha=4.2\times 10^8$,  $\xi= 10000$;\,\,\,
{\bf (c1):} $\alpha=2.2\times 10^8$, 
$\xi= 3000$;\,\,
{\bf (c2):} $\alpha=2.2\times 10^8$,  $\xi=10000$.\,\,\,
Increasing $\xi$ for $\alpha$ fixed (e.g. in plots (b)) 
brings deeper valleys  and compensates the opposite effect of $\lambda$, 
while keeping the height of $\hat W$ fixed; $\xi$ also controls the position 
of the valley, relative to plot (a1). 
Decreasing $\alpha$, for $\xi$ fixed (plots (c) versus (b)),  increases the height of $\hat W$. 
Finally, increasing $\lambda$ to $\lambda=0.6$ in the plots above (other than  (a1)) simply makes
their valleys shallower, without changing the rest. The values of the potential are colour encoded 
with blue (red) for the lowest  (highest) value, respectively. 
}\label{plots}
\end{figure}

As apparent in the plots of Figure~\ref{plots} where quantum corrections are included,
  increasing $\xi$ (and $\tilde\xi$) brings  deeper valleys and also changes their 
initial direction (relative to plot (a1)),
while a smaller $\alpha$ ($\tilde\alpha$) or larger $\tilde M$  increases
the height of the potential in its central region (of small $\phi$). 
The actual inflaton roll is  along a trajectory in the plane ($\phi,\rho$)
that is controlled by: the height of the potential (influenced by $\alpha$), the position/depth
 of the valleys (influenced by $\xi$), by  $H$ and by the field-space metric 
since $\gamma_{\phi\phi}=\exp(-\sqrt{2/3}\,\rho/M_p)$.
The fields then oscillate 
about the  global minimum at $(\phi,\rho)=(0,0)$, where reheating  takes place 
after inflation ends. 

\begin{figure}[ht]
\begin{center}
\includegraphics[width=6.cm,height=5cm]{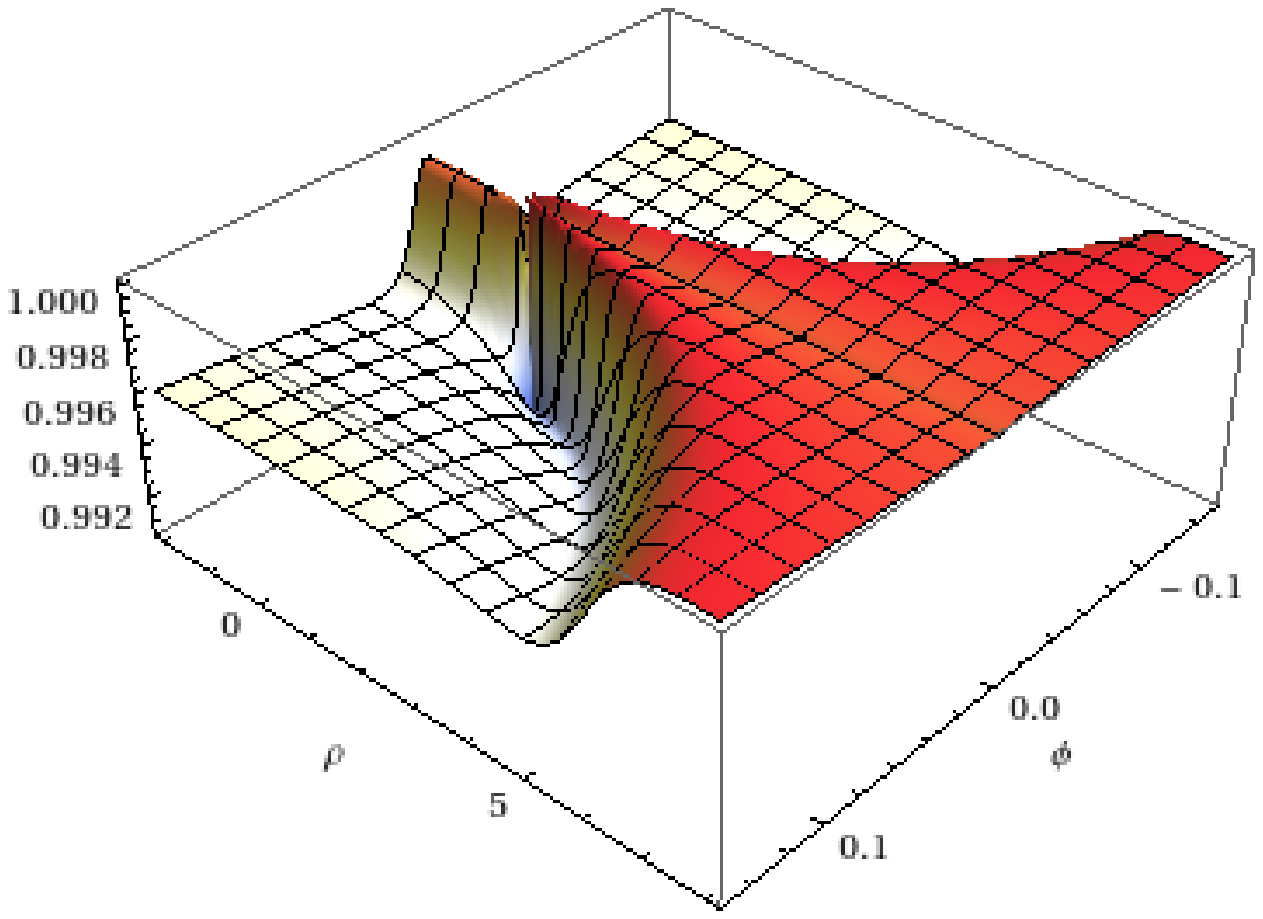}
\includegraphics[width=6.cm,height=5cm]{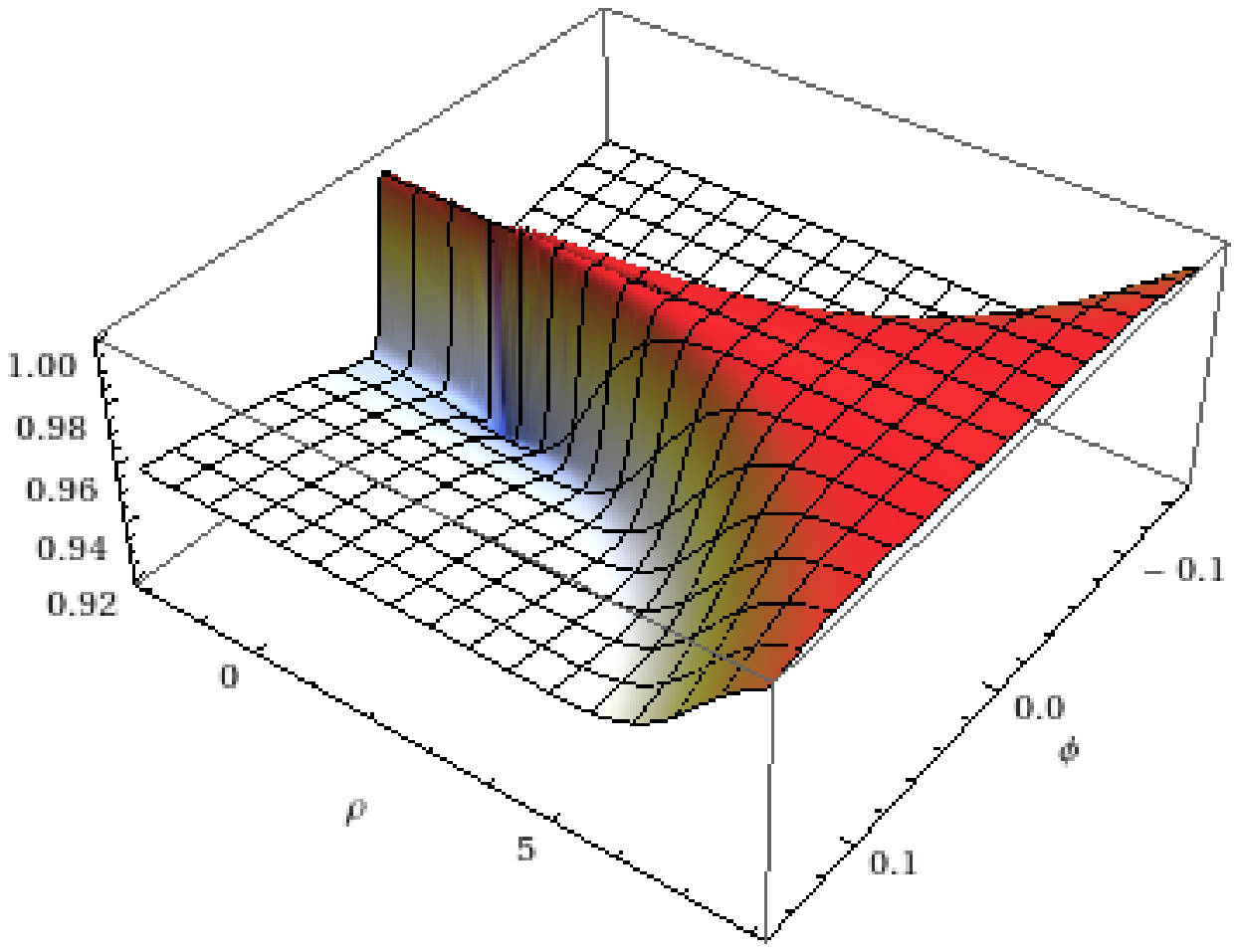}
\end{center}
\caption{\small The relevance of  the quantum corrections to the potential
in terms of the field values in Planck units. 
The plots show the two-loop corrected potential  $\hat W$ of eq.(\ref{f2})
normalized to its classical value eq.(\ref{EE}), for various field values, 
with $\xi= 3000$, $\alpha=4.2\times 10^8$, 
and  $\lambda=0.06$ (left) and $\lambda=0.6$ (right).
These plots show that the relative difference quantum effects bring to the classical
potential
is small, up to $\varepsilon=1\%$ reduction for $\lambda=0.06$, 
 $\varepsilon=6\%$ reduction for $\lambda=0.6$ or 
up to  $\varepsilon=10\%$ for $\lambda=1$ (not shown); 
(note our normalisation for  $\lambda$,  eq.(\ref{L})).
}
\label{fig2}
\end{figure}

Figure~\ref{plots}  shows a behaviour of the potential at the quantum level
that is similar to the classical one analysed recently in \cite{AS}.
The difference between the 
classical picture \cite{AS} and our quantum picture is  usually small,
see  Figure~\ref{fig2} for details.
In this figure it was shown that  the
quantum-corrected $\hat W$  is reduced relative to its classical value, 
by  up to  $6\%$   for $\lambda=0.6$,  and up to
 $10\%$ for $\lambda=1$  (with our normalisation of $\lambda$).
In all these figures, the subtraction scale  was set $\mu=M_p$.  
These results change mildly e.g. by $5\%$ when reducing $\mu$ by a factor of $100$
 at $\lambda=0.6$  (due to a $\log\mu$ dependence of the quantum corrections). As a result, 
the classical level analysis in \cite{AS} 
of the  Starobinsky-Higgs model does not  change significantly
at the quantum level, for the generic values of the parameters used here.

With $\hat W$ mildly  reduced relative to its classical value, 
by one- and two-loop  corrections,  then  $\eta_{\parallel}$ 
increases by a similar relative amount (denoted $\varepsilon$),  see eqs.(\ref{H2}) to (\ref{param})
(we ignore here the subleading two-loop wavefunction correction).
Accordingly, one has  
$\varepsilon \approx 1$\% for $\lambda=0.06$, and $\varepsilon \approx 6$\%  for $\lambda=0.6$.
Therefore, the spectral index $n_s=1+2\eta_{\parallel}-4\epsilon$ \cite{AS} 
decreases by $2\,\varepsilon\, \vert \eta_{\parallel}\vert$. 
For a  conservative $\vert \eta_\parallel\vert\sim 0.05$, the decrease of $n_s$ is 
then between $0.002$ and $0.006$ (depending on $\varepsilon$), which 
is situated between $1\sigma$ and $3\sigma$  deviation of $n_s$  \cite{Planck}
(ignoring a subleading  $\epsilon$ correction relative to $\eta_\parallel$).
For the tensor-to-scalar ratio  $r=16\,\epsilon \, c_s$ \cite{AS}
a similar calculation gives a relative increase 
of $2.5\, \varepsilon$ from its classical value, or $2.5\%$ ($13\%$) for $\lambda=0.06$ ($\lambda=0.6$),
respectively ($n_s$ and $r$ variations increase with  $\lambda$).

\section{Conclusions}

In this work we analysed the two-loop matter corrections (of Higgs-like $\phi$) to the action
in curved space-time, in the presence of  non-minimal coupling $\xi\,\phi^2\,R$ and of 
terms quadratic in the curvature scalar ($R$) and tensors ($R_{\mu\nu}$, $R_{\mu\nu\rho\sigma}$). 
The motivation was to explore the effects of the quantum
corrections in classical models of inflation, such as Starobinsky-Higgs  inflation.
The scalar potential is then an expansion $V(\phi)=V_0+V_1+ V_2+\cdots$,  where $V_0$ is the flat 
space-time potential, $V_1$  is linear in $R$,  $V_2$ contains is quadratic in 
$R$, $R_{\mu\nu}$,  $R_{\mu\nu\rho\sigma}$, etc, with coefficients functions of $\phi$.
The quantum corrected  $V(\phi)$ was computed  from  the corresponding 
Callan-Symanzik equations for  $V_{0,1,2}$ at the two-loop level.
The  quantum potential is
 useful for precision data constraints on the Starobinsky-Higgs model.

While an $R^2$-term may be absent classically, it is nevertheless present at the quantum 
level with a $\phi$-dependent  coefficient. 
Given the presence of this term, the 
action automatically includes a Starobinsky-like model as a limiting case.
Correspondingly, one has a new  scalar field (scalaron $\rho$) induced geometrically 
by the $R^2$-term. As a result,  in the presence of non-minimal coupling
multi-field inflation is  a quantum  consequence.

Depending on the field and parameters values
one can have dominant Higgs, dominant Starobinsky  inflation, or a
combination of these,  thus giving a unified (quantum) model of Starobinsky-Higgs inflation.
The quantum corrections (sum of one- and two-loop) to the potential
are  of few percent order (up to $10\%$ if $\lambda=1$)
 and they increase with  $\lambda$; 
their impact on the spectral index $n_s$ is between 
$1\sigma$ and $3\sigma$ variations, while $r$ increases by $2.5\%$ to 
$13\%$ relative to its classical value (depending on exact $\lambda$).
Thus,  the Starobinsky-Higgs model of  inflation is 
rather stable in the  presence of quantum corrections, for generic $\lambda$.

It is known that for a large non-minimal coupling $\xi$, one can integrate 
the Higgs field  in the classical action to recover the ``usual'' Starobinsky model  with 
a modified  scalaron mass  that depends on all parameters:  $\alpha$, $\xi$, $\lambda$. 
At the quantum level, we showed that integrating $\phi$ generates a 
``refined''  Starobinsky-like  action with additional terms
$\xi^2\,R^2  \ln^n (\xi \vert R\vert/\lambda\mu^2)$, $n=1,2$, with implications 
for inflation (spectral index and scalar-to-tensor ratio) discussed recently.
These terms bring corrections to the scalaron potential
$\propto\lambda\,\rho/M_p$ (Einstein frame) and a ``running'' scalaron mass.
Terms like  $C^2 \ln(\xi\vert R\vert/\mu^2)$ and $G\ln(\xi\vert R\vert/\mu^2)$ 
were also  generated when integrating matter field(s) ($\phi$) at the loop level.

Interestingly, in the case  the $R^2$-term is absent at the classical level, we showed that 
the  ``usual'' Starobinsky inflation can still take place. Quantum corrections
generate a leading $\lambda$-independent
 term  $\xi^2 R^2\ln (M_p^2/\phi^2)$, with higher orders suppressed by $\lambda$.  
 For  suitable  $\xi$ and a small (fixed) Higgs field  $\phi^2\ll M_p^2/\xi$,  the ``usual'' 
Starobinsky  inflation is therefore generated at  the quantum  level alone, which  is another
result of this work. In this case  the Higgs field 
plays  no role other than having a suitable (large)  non-minimal  coupling to $R$.

\vspace{0.1cm}

\begin{center}
------------------------
\end{center}

\vspace{-0.5cm}

\section*{Appendix}

\def\theequation{A-\arabic{equation}}
\def\thesubsection{A}
\setcounter{equation}{0}
\def\thefigure{A-\arabic{figure}}

\subsection*{A: Details on two-loop potential}

$\bullet$ The two-loop potential is found from the Callan-Symanzik general equation
\bea
\Big(\mu\frac{\partial}{\partial \mu}-\gamma\,\phi\, \frac{\partial}{\partial\phi}
+\beta_\lambda\,\frac{\partial}{\partial\lambda}
+\beta_\xi\,\frac{\partial}{\partial\xi}
+\beta_\Lambda\frac{\partial}{\partial \Lambda} 
+\beta_{\kappa_0}\frac{\partial}{\partial \kappa_0}
+\sum_{j=1}^3 \beta_{\alpha_j} \frac{\partial}{\partial \alpha_j}\Big) V=0.
\eea
using an expansion in curvature as mentioned in the text, see also 
eqs.(\ref{L}), (\ref{op}):
\bea
V=V_0+V_1+V_2
\eea
For our  massless case, $\beta_\Lambda=\beta_{\kappa_0}=0$ ($\kappa_0$ is the coefficient of $R$). 
The metric is not 
quantised. $V$ depends on $\alpha_i$ only via  classical action (\ref{L}).
Then
\bea\label{c}
\Big(\mu\frac{\partial}{\partial \mu}
+\beta_\lambda\,\frac{\partial}{\partial\lambda}+\beta_\xi \frac{\partial}{\partial \xi}
-\gamma\,\phi\, \frac{\partial}{\partial\phi}
\Big)\,V= 
\beta_{\alpha_1}\,R^{\mu\nu\rho\sigma}\,R_{\mu\nu\rho\sigma}
+\beta_{\alpha_2}\,R^{\mu\nu} R_{\mu\nu} +\beta_{\alpha_3}\,R^2
\eea
where
\be\label{eq}
\beta_X=\frac{d X}{d\ln \mu}, \,\, X:\lambda, \xi, \alpha_{j};\qquad
\gamma=-\frac{d\ln\phi}{d\ln\mu}=
\frac{d \ln Z^{1/2}_{\phi}}{d\ln\mu};\,\,\,\,
\phi_B=Z_\phi^{1/2}\phi\,\mu^{-\epsilon}
\ee

\medskip\noindent
Equating terms of similar curvature in (\ref{c})
gives the CS equations for $V_0$, $V_1$, $V_2$. For $V_0$ 
\medskip
 \bea
 \Big(\mu\frac{\partial}{\partial \mu} 
+\beta_\lambda\frac{\partial}{\partial \lambda} 
-\gamma\, \phi\frac{\partial}{\partial\phi}\Big) \, V_0=0,
\quad V_0^{(4)}\Big\vert_{\phi=\mu}=\lambda.
 \eea

\medskip\noindent
where  $V_0^{(4)}=d^4 V_0/d\phi^4$. To solve it, first 
differentiate it four times  wrt $\phi$:
\medskip
\bea\label{eq1}
\Big(\mu \frac{\partial}{\partial \mu}
+\beta_\lambda\frac{\partial}{\partial \lambda}
-4\gamma-\gamma\,\phi\frac{\partial}{\partial\phi}\Big)\,V_0^{(4)}=0 
\eea
Since  $V_0^{(4)}$ is dimensionless,  it must depend on $\mu$ and $\phi$ 
only via their ratio. Therefore
\medskip
\bea\label{cs1}
\Big(
\frac{\partial}{\partial t} 
+\tilde\beta_\lambda\frac{\partial}{\partial\lambda}
- 4\tilde\gamma\Big)\, V_0^{(4)}=0,\qquad t=\ln (\mu/\phi),
\qquad  V_0^{(4)}\Big\vert_{t=0}=\lambda
\eea

\medskip\noindent
where
\bea\label{tildebeta}
\tilde\beta_\lambda&=&\frac{\beta_\lambda}{1+\gamma}=\beta_\lambda^{(1)}+\beta_\lambda^{(2)}+\cO(\lambda^4)
\nonumber\\
\quad
 \tilde\gamma&=& \frac{\gamma}{1+\gamma}=\gamma^{(2)}+\cO(\lambda^4).
\eea
The superscripts on $\beta^{(k)}_\lambda$ and $\gamma^{(k)}$ denote the $k$-th loop order correction:
$\beta_\lambda^{(1)}\sim\lambda^2$, $\beta_\lambda^{(2)}\sim\lambda^3$, $\gamma^{(2)}\sim \lambda^2$, etc.
Writing $V_0^{(4)}$ as an expansion in powers of $\lambda$ up to three loop corrections
\medskip
\bea
V_0^{(4)}=\cV^{(0)}+\cV^{(1)}+\cV^{(2)}+\cO(\lambda^4),
\qquad \cV^{(j)}\sim \lambda^{j+1},\,\, j=0,1,2.
\eea

\medskip\noindent
then  $\cV^{(0)}=\lambda$. Then in order $\lambda^2$ the CS equation gives
\medskip
\bea
\frac{\partial \cV^{(1)}}{\partial t}+\beta_\lambda^{(1)} \frac{\partial \cV^{(0)}}{\partial\lambda}
=\cO(\lambda^3),
\eea
with a solution $\cV^{(1)}=-\beta_\lambda^{(1)} \,t+$constant. 
Next
\medskip
\bea
\frac{\partial \cV^{(2)}}{\partial t}+\beta_\lambda^{(2)}\frac{\partial \cV^{(0)}}{\partial\lambda}
+\beta_\lambda^{(1)} \frac{\partial \cV^{(1)}}{\partial \lambda}
-4\gamma^{(2)} \cV^{(0)}=\cO(\lambda^4)
\eea

\medskip\noindent
which is integrated wrt $t$.  We find
\bea\label{dv4}
V_0^{(4)}=
\lambda-\beta_\lambda^{(1)}\,t +\big(4\gamma^{(2)}\lambda-\beta_\lambda^{(2)}\big)\,t
+\beta_\lambda^{(1)} \frac{\partial\beta_\lambda^{(1)}}{\partial \lambda}\frac{t^2}{2}.
\eea

\medskip\noindent
In (\ref{dv4}) denote by $u_1$ and $u_2$ the coefficients of $t$ and $t^2$.
Integrate (\ref{dv4}) over $\phi$ four times, with $t$ as in (\ref{cs1}), then
\medskip
\bea\label{vflat}
\quad V_0
&=&\Big[\frac{\lambda}{4!}+
\frac{25}{288} \,u_1 +\frac{415}{1728}\, u_2\Big]\phi^4 -\frac{1}{288} (6\, u_1+25\, u_2)\, \phi^4\,
\ln\frac{\phi^2}{\mu^2}
+\frac{u_2}{96}\,\phi^4\, \ln^2\frac{\phi^2}{\mu^2}
\nonumber\\
&=&\frac{\lambda}{4!}\,\phi^4
+\frac{\lambda^2}{16\kappa}\, \phi^4\, \Big[\,\ln\frac{\phi^2}{\mu^2}-\frac{25}{6}\,\Big]
+\frac{\lambda^3}{32\kappa^2}\,\phi^4\,\Big[\,\frac{515}{6}
-29\ln\frac{\phi^2}{\mu^2}+3\ln^2\frac{\phi^2}{\mu^2}\,\Big],
\qquad\quad
\eea

\bigskip\noindent
quoted in the text; in the last step we  used (Appendix~B):
\medskip
\bea
\beta_\lambda^{(1)}=\frac{3\lambda^2}{\kappa},\qquad 
\beta_\lambda^{(2)}=-\frac{17}{3}\frac{\lambda^3}{\kappa^2},\qquad
\gamma^{(2)}=\frac{\lambda^2}{12\kappa^2}.
\eea


\bigskip
\medskip\noindent
$\bullet$ Further, the CS equation for $V_1$ is found from (\ref{c})
\bea\label{rr}
\Big( \mu \frac{\partial}{\partial \mu}
+\beta_\lambda\frac{\partial}{\partial\lambda}
+\beta_\xi\frac{\partial}{\partial\xi}
- \gamma\,\phi\frac{\partial}{\partial\phi}
\Big) \,V_1=0.
\eea
with
\bea
 V_1^{(2)}=\frac{\partial^2  V_1}{\partial\phi^2}, \qquad
V_1^{(2)}\Big\vert_{\phi=\mu}=\xi\,R.
\eea
After applying $\partial^2/\partial\phi^2$ one has
\bea\label{cs2}
\Big(
\frac{\partial}{\partial t}
+\tilde\beta_\lambda\frac{\partial}{\partial\lambda}
+\tilde\beta_\xi \,\frac{\partial}{\partial\xi}
- 2\tilde \gamma\Big)\,
V_1^{(2)}=0,
\qquad
 V_1^{(2)}\Big\vert_{t=0}=\xi\,R,
\eea
with
\bea\label{tildebetaxi}
\tilde\beta_\xi=\frac{\beta_\xi}{1+\gamma}=\beta_\xi^{(1)}
+\beta_\xi^{(2)}-\gamma^{(2)}\,\beta_\xi^{(1)}
+\cO(\lambda^4),\qquad\beta_\xi^{(j)}\sim \lambda^j.
\eea

\medskip\noindent
The solution has a loop expansion
\medskip
\bea
V_1^{(2)}=W^{(0)}+W^{(1)}+W^{(2)}+\cO(\lambda^3),\qquad  W^{(j)}\sim \lambda^j,
\,\, j=0,1,2,3.
\eea

\medskip\noindent
Then  $W^{(0)}=\xi R$. The
CS equation then splits into 2 equations for the remaining $W^{(j)}$
\medskip
\bea
&&\frac{\partial W^{(1)}}{\partial t}+\beta_\xi^{(1)} \,\frac{\partial W^{(0)}}{\partial\xi}=
\cO(\lambda^2)
\nonumber\\
&&\frac{\partial W^{(2)}}{\partial t} +\beta_\lambda^{(1)}\frac{\partial W^{(1)}}{\partial\lambda}
+\beta_\xi^{(1)}\frac{\partial W^{(1)}}{\partial\xi}+\beta_\xi^{(2)}\frac{\partial W^{(0)}}{\partial\xi}
-2\gamma^{(2)} \,W^{(0)}=\cO(\lambda^3)
\eea

\medskip\noindent
Then $W^{(1)}=-\beta_\xi^{(1)}\,R\, t$ which is used to find $W^{(2)}$. Adding these together, then
\medskip
\bea\label{qq2}
V_1^{(2)}=R\,\Big[ 
\xi-\beta_\xi^{(1)} \,t +
\Big(\beta_\lambda^{(1)}\,\frac{\partial}{\partial \lambda} \beta_\xi^{(1)}
+\beta_\xi^{(1)}\frac{\partial}{\partial\xi} \beta_\xi^{(1)} \Big)\,\frac{t^2}{2} +
\big(2\gamma^{(2)} \xi-\beta_\xi^{(2)}\big)\,t\,\Big].
\eea

\medskip\noindent
In (\ref{qq2}) denote by $z_1$, $z_2$ the coefficients of $t$ and $t^2$ and integrate
twice over $\phi$, then
\medskip
\bea\label{vone}
V_1&=&R\,\phi^2
\Big\{ \frac{\xi}{2}+\frac34 z_1+\frac74 z_2
-
\frac14  (z_1+3 z_2)\ln\frac{\phi^2}{\mu^2} 
+
\frac18 z_2 \ln^2 \frac{\phi^2}{\mu^2}
\Big\}
\nonumber\\
\!\!&=&\!\!
R\,\phi^2\, \Big\{\,
\frac{\xi}{2} + \frac{\lambda}{4\,\kappa} \,\Big(\xi \p \frac{1}{6}\Big)
\Big(\ln\frac{\phi^2}{\mu^2}-3\Big) 
-\frac{\lambda^2}{4\,\kappa^2}\,\Big(\frac{7}{6}\,\xi \p  \frac{13}{36}\Big)
 \,\Big(\ln\frac{\phi^2}{\mu^2}-3\Big) 
\nonumber\\
&+&\!\!\!\!
\frac{2\lambda^2}{\kappa^2} \frac{1}{4} \Big(\xi \p  \frac{1}{6}\Big) 
\Big(7-3\ln\frac{\phi^2}{\mu^2} +\frac{1}{2}\ln^2 \frac{\phi^2}{\mu^2} \Big)
\Big\},
\eea

 \medskip\noindent
where in the last step we replaced the beta functions (see Appendix B)
\medskip
\bea
\beta_\xi^{(1)}=\frac{\lambda}{\kappa} \Big(\xi\p \frac{1}{6}\Big),
\quad
\beta_\xi^{(2)}=-\frac{\lambda^2}{\kappa^2} \Big(\xi\p \frac{1}{6}\Big) 
\m \frac{\lambda^2}{\kappa^2}\,\frac{7}{36}
\eea

\medskip\noindent
($\xi=\m 1/6$ is conformal at one-loop only).


\vspace{1.cm}
\noindent
$\bullet$ Finally, compute the two-loop corrections to $V_2$
\medskip
\bea\label{vtwo}
V_2\!\!\!&=&\!\! 
-\tilde \alpha_1 \,R_{\mu\nu\rho\sigma} R^{\mu\nu\rho\sigma}
-\tilde \alpha_2 \,R_{\mu\nu} R^{\mu\nu}
-\tilde\alpha_3 \,R^2
\eea

\medskip\noindent
The  CS equation is
\medskip
\bea
\Big(
\mu\frac{\partial}{\partial \mu}
+\beta_\lambda\,\frac{\partial}{\partial\lambda} 
+\beta_\xi\,\frac{\partial}{\partial\xi}
-\gamma\phi\frac{\partial}{\partial\phi}\Big)\,
(-\tilde \alpha_i)
=
\beta_{\alpha_i},\qquad i=1,2,3.
\eea
giving
\bea\label{cs3}
\Big(
\frac{\partial}{\partial t} 
+\tilde\beta_\lambda  \frac{\partial}{\partial\lambda}
+\tilde\beta_\xi \frac{\partial}{\partial\xi} \Big)\,
(-\tilde\alpha_i)
=   \tilde\beta_{\alpha_i},
\qquad
\tilde\alpha_i\Big\vert_{t=0}=
\alpha_i.
\eea

\medskip\noindent
Here $\tilde\beta_\lambda$ and $\tilde\beta_\xi$ are 
given in eqs.(\ref{tildebeta}) and (\ref{tildebetaxi})
and
\medskip
\bea
\tilde\beta_{\alpha_i}
=\frac{\beta_{\alpha_i}}{1+\gamma}
=\beta_{\alpha_i} (1 -\gamma^{(2)}-\gamma^{(3)})+\cO(\lambda^4),
\eea

\medskip\noindent
with $\beta_{\alpha_i}=d\alpha_i/d(\ln\mu)$. A solution is expanded in powers of $\lambda$:
\medskip
\bea
\tilde\alpha_i=\alpha_i^{(0)}+\alpha_i^{(1)}+\cdots, \qquad \alpha_i^{(k)}\sim \lambda^k.
\eea
giving
\bea
- \frac{\partial \alpha_i^{(0)}}{\partial t}=  \beta_{\alpha_i}^{(1)}, 
\qquad\quad
- \frac{\partial \alpha_i^{(1)}}{\partial t}
- \beta_\xi^{(1)}\,\frac{\partial \alpha_i^{(0)}}{\partial \xi}
= \beta_{\alpha_i}^{(2)}.
\eea

\medskip\noindent
Here $\beta_{\alpha_i}^{(1)}\!\sim\!\lambda^0$ and $\beta_{\alpha_i}^{(2)}\!\sim\!\lambda$
are one- and two-loop corrections to $\beta_{\alpha_i}$.
Then at two-loop
\medskip
\bea
-\tilde\alpha_i=
-\alpha_i+\beta_{\alpha_i}^{(1)}\,t
-\beta_\xi^{(1)}\frac{\partial  \beta_{\alpha_i}^{(1)}}{\partial\xi} 
\,\frac{t^2}{2}+\beta_{\alpha_i}^{(2)}
\,t.
\eea

\medskip\noindent
giving,  with $t=\ln \mu/\phi$:
\bea\label{vtwop}
&&
-\tilde\alpha_1
= -\alpha_1+\Big(\frac{-1}{180\kappa}\Big)\,\ln\frac{\mu}{\phi}
\nonumber\\
&& 
-\tilde\alpha_2
=-\alpha_2+\Big(\frac{1}{180\kappa}\Big)\,\ln\frac{\mu}{\phi}
\nonumber\\
&&
-\tilde\alpha_3
=-\alpha_3-\frac{1}{2\,\kappa} \Big(\xi\p \frac{1}{6}\Big)^2 \ln\frac{\mu}{\phi}
+\frac{\lambda}{2\,\kappa^2}\,\Big(\xi\p\frac16\Big)^2\ln^2 \frac{\mu}{\phi},
\eea

\medskip\noindent
quoted in the text. Above we used  (Appendix~B)
\medskip
\be
 \beta_{\alpha_1}^{(1)}=-\frac{1}{180\kappa},\quad
 \beta_{\alpha_2}^{(1)}=\frac{1}{180\kappa}, \quad 
 \beta_{\alpha_3}^{(1)}=-\frac{1}{2\,\kappa} \Big(\xi  \p \frac16\Big)^2,
\quad
 \beta_{\alpha_1}^{(2)}=\beta_{\alpha_2}^{(2)}=\beta_{\alpha_3}^{(2)}=0.
\ee

\vspace{0.1cm}
\subsection*{B: Two-loop beta functions}\label{AppendixB}

\def\theequation{B-\arabic{equation}}
\def\thesubsection{B}
\setcounter{equation}{0}
\def\thefigure{B-\arabic{figure}}

Here we provide the derivation of the beta functions $\beta_\xi$ and $\beta_{\alpha_j}$
used in previous section.
With initial $L$ as in (\ref{L}), these  are found from
the two-loop counterterms $\delta L$ to $L$   \cite{PT}. $\delta L$
is of the same form as the terms in initial $L$ eq.(\ref{L}):
\medskip
\be
\frac{\delta L}{\sqrt{\tilde g}}
=-\frac12 \delta Z_\phi\,\phi\Box\phi
-\frac12 \delta\xi\,R\,\phi^2 -\frac{1}{4!}\,\delta\lambda\,\phi^4
+\delta\alpha_1\,R_{\mu\nu\rho\sigma}R^{\mu\nu\rho\sigma}
+\delta\alpha_2 \,R_{\mu\nu}\,R^{\mu\nu}
+\delta\alpha_3\,R^2
\ee

\medskip\noindent
a). First, $\delta Z_\phi=Z_\phi-1$ and
\bea
\phi_B=\mu^{-\epsilon}\,Z_\phi^{1/2}\,\phi
\eea
with
\bea
Z_\phi=1-\frac{\lambda^2}{24\kappa^2\,\epsilon}, \qquad 
\gamma_\phi^{(2)}=\frac{d\ln Z_\phi^{1/2}}{d\ln\mu}=\frac{\lambda^2}{12\kappa^2}.
\eea

\medskip\noindent
where we used $\beta_\lambda$ derived below:
\smallskip
\bea
\lambda_B= \mu^{2\epsilon}\,(\lambda+\delta\lambda)\,Z_\phi^{-2},\qquad
\delta\lambda\equiv \lambda \,(Z_\lambda-1)=\delta\lambda^{(1)}+\delta\lambda^{(2)}
\eea
with
\bea
\delta\lambda^{(1)}=\frac{-3\lambda^2}{(-2\epsilon)\,\kappa},
\qquad
 \delta\lambda^{(2)}=\frac{\lambda^3}{\kappa^2}\,\Big(\frac{9}{4\epsilon^2}-\frac{3}{2\epsilon}\Big)
\eea

\medskip\noindent
in the MS scheme. Since $\frac{d \lambda_B}{d\ln\mu}=0$, with
$Z_\phi$ above and $\beta_\lambda=d\lambda/d(\ln\mu)$,
then at two-loop
\bea
\beta_\lambda=\beta_\lambda^{(1)}+\beta_\lambda^{(2)}=-2\epsilon\lambda+\frac{3\lambda^2}{\kappa}-\frac{17}{3}\frac{\lambda^3}{\kappa^2}.
\eea
b).  Further, we have:
\medskip
\bea
&& \xi_B=(\xi+\delta\xi)\,Z_\phi^{-1},\qquad \delta\xi\equiv \xi\,(Z_\xi-1)
=\delta\xi^{(1)}+\delta\xi^{(2)}\qquad\qquad
\eea
with \cite{PT}
\bea\label{por}
\delta\xi^{(1)}=\frac{-\lambda}{\kappa\,(-2\epsilon)} \Big(\xi \p  \frac16\Big),
\qquad
 \delta\xi^{(2)}=\frac{\lambda^2}{2\kappa^2\,\epsilon^2} \Big(\xi \p  \frac16\Big)
-\frac{\lambda^2}{4\kappa^2\,\epsilon} \Big(\xi \p  \frac{7}{36}\Big)
\eea

\medskip\noindent
Together with $Z_\phi$ and  $\xi_B=\xi\,Z_\xi Z_\phi^{-1}$, (\ref{por})
gives the two-loop corrected $\beta_\xi=d\xi/d(\ln\mu)$ as
\medskip
\bea
\beta_\xi= \beta_\xi^{(1)}+\beta_\xi^{(2)}=
\frac{\lambda}{\kappa} \Big(\xi\p  \frac16\Big)
+
\frac{-\lambda^2}{\kappa^2}\Big(\xi\p \frac16\Big)\m  \frac{7}{36}\frac{\lambda^2}{\kappa^2}.
\eea
c). Finally one has
\bea
&& \alpha_{j B}=\mu^{-2\epsilon}\,(\alpha_j+\delta\alpha_j),\quad 
\delta\alpha_j\equiv \alpha_j\,(Z_{\alpha_j}-1)=\delta\alpha_j^{(1)}+\delta\alpha_j^{(2)},\quad j:1,2,3.
\eea
with \cite{PT}
\bea
\delta\alpha_1^{(1)}=\frac{1}{(-2\epsilon)\,\kappa}\frac{1}{180}=-\delta\alpha_2^{(1)},
\qquad\delta\alpha_1^{(2)}=\delta\alpha_2^{(2)}=0,
\eea
and
\bea
\delta\alpha_3^{(1)}=\frac{1}{2\,(-2\epsilon)\,\kappa}
\Big(\xi \p  \frac16\Big)^2,
\qquad
\delta\alpha_3^{(2)}=-\frac{\lambda}{2\kappa^2\,(2\epsilon)^2} \Big(\xi  \p  \frac16\Big)^2
\eea

\medskip\noindent
With  $\beta_{\alpha_j}=d\alpha_j/d (\ln\mu)$ from  $d\alpha_{j\, B}/d\ln\mu=0$ and
$\alpha_{j\,B}=\mu^{-2\epsilon} \alpha_j\,Z_{\alpha_j}$,  one finds 
\bea
\beta_{\alpha_j}=\beta_{\alpha_j}^{(1)}+\beta_{\alpha_j}^{(2)}
\eea
where
\bea
&&\beta_{\alpha_1}^{(1)}=2\epsilon\,\alpha_1-\frac{1}{180\,\kappa},
\nonumber\\
&&\beta_{\alpha_2}^{(1)}=2\epsilon\,\alpha_2+\frac{1}{180\,\kappa},
\nonumber\\
&&\beta_{\alpha_3}^{(1)}=2\epsilon\,\alpha_3-\frac{1}{2\,\kappa}
\Big( \xi\p \frac16 \Big)^2,
\eea
while $\beta_{\alpha_{1,2,3}}^{(2)}=0$ (as expected).

\bigskip
\bigskip\noindent
{\bf Acknowledgements:  }
The author thanks Hyun Min Lee (Seoul, Chung-Ang University) and Andrei Micu (IFIN Bucharest)
for  interesting  discussions on  quantum corrections in two-field inflation models.

\end{document}